\documentclass[twocolumn]{aastex631}
\usepackage{amsmath}
\usepackage{fontawesome}

\usepackage{algorithmic}
\received{February 28, 2026}
\revised{May 6, 2026}
\accepted{May 8, 2026}
\shorttitle{Bayesian Doppler Imaging}
\shortauthors{Ureshino et al.}
\begin{document}
%\linenumbers
\title{Bayesian Doppler Imaging: Simultaneous Inference of Surface Maps and Geometric Parameters}

\correspondingauthor{Hajime Kawahara} 

\author{Yamato Ureshino}
%\email{yamatonmail@g.ecc.u-tokyo.ac.jp}
\affiliation{Department of Space Astronomy and Astrophysics, ISAS/JAXA, 3-1-1, Yoshinodai, Sagamihara, Kanagawa, 252-5210 Japan}
\affiliation{Department of Astronomy, Graduate School of Science, The University of Tokyo, 7-3-1 Hongo, Bunkyo-ku, Tokyo 113-0033, Japan}

\author[0000-0003-3309-9134]{Hajime Kawahara}
\email{kawahara@ir.isas.jaxa.jp}
\affiliation{Department of Space Astronomy and Astrophysics, ISAS/JAXA, 3-1-1, Yoshinodai, Sagamihara, Kanagawa, 252-5210 Japan}
\affiliation{Department of Astronomy, Graduate School of Science, The University of Tokyo, 7-3-1 Hongo, Bunkyo-ku, Tokyo 113-0033, Japan}

\author[0000-0001-7692-0581]{Hibiki Yama}
%\email{yama@iral.ess.sci.osaka-u.ac.jp}
\affiliation{Department of Earth and Space Science, Osaka University, Osaka 560-0043, Japan}

\author[0000-0003-1298-9699]{Kento Masuda}
%\email{kmasuda@ess.sci.osaka-u.ac.jp}
\affiliation{Department of Earth and Space Science, Graduate School of Science, Osaka University, 1-1 Machikaneyama-cho, Toyonaka, Osaka 560-0043, Japan}

\nocollaboration{4}

\begin{abstract}
We present a fully Bayesian, pixel-based Doppler imaging framework that enables the simultaneous inference of surface brightness maps and geometric parameters, including the inclination $i$ and equatorial rotation velocity $v_{\mathrm{rot}}$, from high-resolution spectral time series. 
We treat the inference as a Bayesian linear inverse problem conditioned on nonlinear geometric parameters. The surface map is modeled as a Gaussian Process prior over pixel intensities, introducing a characteristic spatial scale that sets the map resolution. This allows analytical marginalization of the linear coefficients and efficient sampling of the nonlinear parameters with Hamiltonian Monte Carlo.
{Validation with synthetic data demonstrates that our method recovers the longitudes of large-scale surface inhomogeneities and constrains $v_{\mathrm{rot}}$ and $i$  under the adopted model assumptions, while also revealing the limited latitudinal sensitivity intrinsic to Doppler imaging.}
We applied this framework to high-resolution VLT/CRIRES observations of the brown dwarf Luhman 16B. Our analysis reveals a large-scale dark region at mid-latitudes, consistent with previous studies but now with spatially resolved uncertainty estimates. Furthermore, we successfully constrained the geometric parameters {without fixing \(v_{\mathrm{rot}}\sin i\) or $i$ to literature values}, deriving an inclination of $i = 61.0_{-12.3}^{+14.3}$ degrees and an equatorial rotation velocity of $v_{\mathrm{rot}} = 31.2_{-3.1}^{+5.3}~\mathrm{km\,s^{-1}}$. These results indicate a radius broadly consistent with evolutionary models and 
{suggest a possible spin-axis misalignment under the assumption of comparable equatorial rotation velocities for the two components.} 
Our code is publicly available under the MIT license\footnote{\cite{yamato_ureshino_2026_20024778} as the frozen version in Zenodo. \\ https://github.com/prvjapan/BayesianDI}.
\end{abstract}


\section{Introduction}
{
Surface mapping of stars, brown dwarfs, and exoplanets relies on techniques that infer spatial structure from time-dependent observables, with the exception of supergiants, which can be directly spatially resolved by interferometry. In exoplanet studies, phase-curve inversions provide longitudinal brightness maps {\citep{2007Natur.447..183K,2008ApJ...678L.129C}}. Early phase and eclipse analyses already indicated non-uniform dayside brightness distributions (e.g., \citealt{2010ApJ...721.1861A}), and more recent observations have enabled detailed phase-curve mapping (e.g., \citealt{2018AJ....156...17K}). Eclipse mapping exploits secondary-eclipse light curves to constrain two-dimensional surface distributions \citep{2012ApJ...747L..20M,2012A&A...548A.128D,2024AJ....168....4H,2025NatAs...9.1821C}. For directly imaged terrestrial exoplanets, spin-orbit mapping has been proposed to recover two-dimensional surface maps from reflected-light photometric modulation \citep{2010ApJ...720.1333K,2011ApJ...739L..62K,2012ApJ...755..101F}.
}

In contrast to photometric techniques, Doppler imaging is a spectroscopic method that reconstructs two-dimensional surface maps from time-dependent distortions in spectral line profiles \citep{1958IAUS....6..209D, 1974ApJ...192..409F, 1977SvAL....3..147G, 1976AN....297..203K, 1982SvA....26..690G, Vogt_1983}. Originally developed for stars, it has recently been extended to substellar objects. \citet{2014Natur.505..654C} applied it to the brown dwarf on the L-T transition Luhman 16B, revealing surface inhomogeneity, and {\citet{Chen_2024} later performed multi-band Doppler imaging of the Luhman 16AB system, reporting persistent spot-like structures
on Luhman 16B and polar spots on Luhman 16A.}  Given the widespread photometric variability of brown dwarfs, commonly attributed to large-scale cloud patterns \citep[e.g.][]{2017AstRv..13....1B,2018haex.bookE..94A, Tan_2025}, Doppler imaging offers a means to probe atmospheric structure and dynamics. Motivated by the successful measurement of the spin rotation of $\beta$ Pictoris b \citep{2014Natur.509...63S}, extensions to directly imaged exoplanets with Extremely Large Telescopes have also been proposed \citep{2014A&A...566A.130C}. Recent studies have further evaluated the feasibility of such observations, exploring the capabilities of next-generation ELT spectrographs for high-resolution surface mapping \citep{Plummer_2023, 2025ExA....59...29P}.

Doppler imaging is an ill-posed inverse problem and has traditionally been treated as a point estimation through the regularization such as maximum entropy method \citep[e.g.][]{1987ApJ...321..496V} and Tikhonov regularization \citep[e.g.][]{1990A&A...230..363P}. A fully Bayesian formulation would allow quantitative assessment of the credibility of the inferred maps, but remains computationally challenging because of the high dimensionality of the surface map. In addition, physically important parameters such as inclination ($i$), projected rotational velocity ($v_{\mathrm{rot}} \sin i$), and rotation period ($P$) are often fixed using external constraints, although they should ideally be inferred jointly.

Toward a Bayesian formulation of Doppler imaging, \citet{refId0} proposed an approximate Bayesian approach in which the posterior distribution is modeled using modern machine-learning techniques, specifically normalizing flows, to directly learn an approximation to the posterior. Separately, \citet{2021arXiv211006271L} extended their spherical-harmonic (SH)-based simulator {\sf starry} to enable Doppler imaging, { demonstrating its application to Luhman 16B using the same dataset as \citet{2014Natur.505..654C}. Using a different analytical framework, \citet{Plummer_2022} developed a spot-based Doppler imaging technique and similarly applied it to the Luhman 16B data to characterize its surface inhomogeneities.} In principle, these frameworks allow Bayesian inference via Markov Chain Monte Carlo (MCMC) sampling. However, a systematic validation of whether posterior distributions for the inclination, $v_{\mathrm{rot}} \sin i$, and rotation period can be robustly recovered remains limited.

A Bayesian formulation has also been developed for spin-orbit mapping, which, like Doppler imaging, involves many map parameters and only a few nonlinear parameters. \citet{2018AJ....156..146F} modeled the pixel-based surface map as a Gaussian process on the sphere and performed MCMC sampling, constituting the first Bayesian treatment of spin-orbit mapping. However, direct MCMC sampling of the high-dimensional map is computationally expensive.
\citet{2020ApJ...900...48K} addressed this by analytically marginalizing the linear map parameters and sampling only the nonlinear parameters with MCMC, followed by conditional draws of the map. This enabled joint posterior inference and demonstrated robust recovery of nonlinear quantities such as obliquity and spatial scale. The same framework has since been applied to spiral structure extraction in ALMA observations of protoplanetary disks \citep{2024MNRAS.532.1361A}.

In this work, we extend this approach to Doppler imaging and develop a fully Bayesian framework to infer the surface map and nonlinear parameters, inclination and rotation velocity, from spectral time series. The method is implemented in {\sf JAX} \citep{jax2018github}, enabling efficient sampling with Hamiltonian Monte Carlo and the No-U-Turn Sampler (HMC-NUTS).

The remainder of this paper is organized as follows. In Section 2, we formulate Doppler imaging as a linear inverse problem conditional on a small set of nonlinear parameters and describe an efficient posterior sampling method. Section 3 validates the framework using synthetic data, recovering injected maps as well as inclination and rotation velocity from spectra alone. In Section 4, we apply the method to the spectral time series of Luhman 16B and infer posterior distributions for the surface map and geometric parameters. Section 5 summarizes and discusses the results.

\section{Bayesian Doppler Imaging}
\subsection{Doppler Imaging as a Linear Inverse Problem}\label{discretizing}

In this section, we formulate Doppler imaging as a linear inverse problem in the surface map. Although its linearity has been recognized since \citet{1987ApJ...321..496V}, we restate it to provide the basis for our method. This formulation enables a direct extension to a Bayesian linear inverse problem, as described in Section \ref{sec:bayes}, where the model is written explicitly in probabilistic form.

\subsubsection{Observation Model}
The forward model for Doppler imaging describes how the surface brightness distribution on a rotating body is mapped into the phase-resolved line profiles observed by a distant observer. In the continuous form, the monochromatic flux at wavelength~$\lambda$ and rotational phase~$\varphi$ can be expressed as an integral over the visible hemisphere:
\begin{align}
  d(\lambda,\varphi)
  &= \int_{\mu>0}\!
  \mu(\theta^\ast,\phi^\ast;\varphi)\,
  L(\theta^\ast,\phi^\ast;\varphi)\,\nonumber\\
  &\phantom{=}~\qquad\qquad
  a(\theta^\ast,\phi^\ast)\,
  s^\ast\!\left(\frac{\lambda}{D(\theta^\ast,\phi^\ast;\varphi)}\right)
  \,d\Omega,
  \label{eq:continuous}
\end{align}
where $(\theta^\ast,\phi^\ast)$ denote the colatitude and longitude on the surface of the object, $a(\theta^\ast,\phi^\ast)$ is the surface brightness distribution, and $s^\ast(\lambda)$ is the intrinsic (rest-frame) line profile. The factor $\mu$ is the direction cosine between the local surface normal and the line of sight (projected area factor).
%, which also selects the visible hemisphere ($\mu>0$). 
The integration is carried out over the visible area, that is, the region where $\mu > 0$. 
The functions $L$ and $D$ represent the limb-darkening weighting and the local Doppler factor, respectively. Their explicit forms will be introduced in \S \ref{construction_W}.

Since Eq.~(\ref{eq:continuous}) is linear with respect to the surface brightness map $a(\theta^\ast,\phi^\ast)$, it can be discretized and written in a matrix form as
\begin{equation}
  \boldsymbol{d} = W\,\boldsymbol{a},
  \label{eq:matrix-form}
\end{equation}
where $\boldsymbol{d}$ and $\boldsymbol{a}$ denote the data vector and the discretized surface map, respectively. The matrix $W$ encapsulates the geometric projection, limb-darkening, and Doppler broadening effects.

\subsubsection{Construction of the design matrix}\label{construction_W}

\begin{figure}[htbp]
  \centering
  \includegraphics[width=\linewidth]{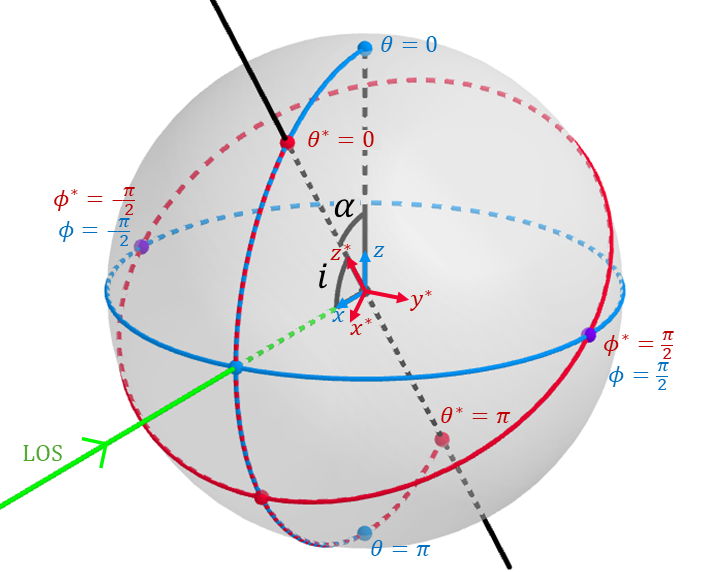}
  \caption{Geometry of the coordinate transformation when the rotational phase is $\varphi_k = 0$.}
  \label{fig1}
\end{figure}

We divide the surface of the object into $N_{\mathrm{pix}}$ equal-area pixels, numbered by $j = 1, 2, \ldots, N_{\mathrm{pix}}$. Each pixel is specified by its colatitude and longitude $(\theta^\ast_j, \phi^\ast_j)$ defined in the co-rotating (stellar) frame, where $\theta^\ast=0$ corresponds to the north pole, $\theta^\ast=\pi$ to the south pole, and $\phi^\ast=0$ to the prime meridian.
We also define the rotational phase $\varphi_k$ ($k = 1, 2, \ldots, N_{\mathrm{phase}}$) such that $\varphi=0$ corresponds to the epoch when the prime meridian faces the observer. At a given phase $\varphi_k$, a surface element originally located at longitude $\phi^\ast_j$ in the co-rotating frame appears at an effective longitude
\begin{align}
  \phi^\ast_{jk} = \phi^\ast_j + \varphi_k.
\end{align}
{The radial velocity of the $j$-th component at the $k$-th phase is given by (see Appendix \ref{app:rv_ele})
\begin{align}
  v_{\mathrm{los}}
%  &= \hat{\boldsymbol{n}}_{\mathrm{obs}}\!\cdot\!\boldsymbol{v} \\
  &= v_{\mathrm{rot}}\cos\alpha\,
     \sin\theta^\ast_j\sin\phi^\ast_{jk}.
\end{align}
} The corresponding Doppler factor affecting the wavelength is
\begin{align}
  D_{jk} = \frac{1+\beta}{\sqrt{1-\beta^2}},
\end{align}
where $\beta$ is the line-of-sight velocity in units of the speed of
light:
\begin{align}
  \beta =
  \frac{v_{\mathrm{los}}(\theta^\ast_j,\phi^\ast_{jk})}{c}.
\end{align}
The spectrum observed from pixel~$j$ at phase~$\varphi_k$, $\boldsymbol{s}_{jk}$, is therefore a Doppler-shifted version of the intrinsic (rest-frame) spectrum $\boldsymbol{s}^\ast$, stretched in wavelength by a factor $D_{jk}$.  To represent both $\boldsymbol{s}_{jk}$ and $\boldsymbol{s}^\ast$ on the same $N_{\mathrm{wav}}$-point wavelength grid, we apply linear interpolation. {Then, we can express the Doppler-shift operation by the linear transform as 
\begin{align}
  \boldsymbol{s}_{jk} = C^{(jk)}\,\boldsymbol{s}^\ast.
\end{align}
The explicit form of $C^{jk}$ is given in Appendix \ref{app:interpolation}.}

The effect of limb darkening is expressed by a coefficient $u$ as
\begin{align}
  L(\mu_{jk}) = 1 - u\,(1 - \mu_{jk}),
\end{align}
where $\mu_{jk}$ is the cosine of the angle $\vartheta_{jk}$ between the local surface normal and the line of sight to the observer:
\begin{align}
  \mu_{jk}
  = \cos\vartheta_{jk}
  = \begin{pmatrix}1\\0\\0\end{pmatrix}\!\cdot\!
    \begin{pmatrix}x\\y\\z\end{pmatrix}
  = \sin\theta_{jk}\cos\phi_{jk}.
\end{align}

The apparent area of each surface element is proportional to $\mu_{jk}$. Only the elements with $\mu_{jk} \ge 0$ are visible, so the geometric weight can be expressed compactly as $\max(0,\mu_{jk})$. Using this weight, the observed spectrum at rotational phase
$\varphi_k$ is constructed as
\begin{align}
  \boldsymbol{d}_k
  &= \sum_{j}
     \max(0,\mu_{jk})\,L(\mu_{jk})\,\boldsymbol{s}_{jk}\,a_j.
\end{align}
Let $W^{(k)}$ denote the matrix obtained by concatenating the column vectors $\max(0,\mu_{jk})\,L(\mu_{jk})\,\boldsymbol{s}_{jk}$ for $j = 1, 2, \ldots, N_{\mathrm{pix}}$. The observed spectrum at phase~$\varphi_k$, which has a length of $N_{\mathrm{wav}}$, can then be written as
\begin{align}
  \boldsymbol{d}_k = W^{(k)}\,\boldsymbol{a}.
\end{align}
Furthermore, by stacking the spectra obtained at all $N_{\mathrm{phase}}$ rotational phases into a single vector of length $N_{\mathrm{wav}}N_{\mathrm{phase}}$,
\begin{align}
  \boldsymbol{d} &=
  \begin{pmatrix}
    \boldsymbol{d}_1\\[3pt]
    \boldsymbol{d}_2\\[3pt]
    \vdots\\[3pt]
    \boldsymbol{d}_{N_{\mathrm{phase}}}
  \end{pmatrix},
\end{align}
and multiplying each block $W^{(k)}$ by a phase-dependent weight $w_k$ that accounts for, e.g., the signal-to-noise ratio or exposure time, the full design matrix $W$ is given by
\begin{align}
  W &=
  \begin{pmatrix}
    w_1\,W^{(1)}\\[3pt]
    w_2\,W^{(2)}\\[3pt]
    \vdots\\[3pt]
    w_{N_{\mathrm{phase}}}\,W^{(N_{\mathrm{phase}})}
  \end{pmatrix}.
\end{align}
The complete linear relation between the data vector and the surface map is therefore expressed as
\begin{align}
  \boldsymbol{d} = W\,\boldsymbol{a}.
  \label{linear}
\end{align}

\subsection{Bayesian inference framework}\label{sec:bayes}

The probabilistic reformulation of linear inverse problems using multivariate Gaussian priors and likelihoods was pioneered by A. Tarantola in seismic tomography \citep{tarantola}. We adopt this framework and model the covariance as a Gaussian process, following \citet{2018AJ....156..146F}, thereby enforcing map smoothness and parameterizing its spatial scale. In this formulation, the linear parameters $\boldsymbol{a}$ can be analytically marginalized, yielding a likelihood that depends only on the nonlinear parameters. This enables MCMC sampling to be restricted to the nonlinear subspace, as proposed by \citet{2020ApJ...900...48K}. In Section \ref{sec:bayes}, we derive the corresponding expressions for Doppler imaging.

\subsubsection{Probabilistic model}

Let $n = N_{\mathrm{phase}}N_{\mathrm{wav}}$ denote the number of data points and $p = N_{\mathrm{pix}}$ the number of surface elements. Building upon the linear relation derived in Eq.~(\ref{linear}), we now include observational noise explicitly and formulate a probabilistic model:
\begin{align}
  \boldsymbol{d}
  &= W\,\boldsymbol{a} + \boldsymbol{\epsilon},\qquad
  \boldsymbol{\epsilon} \sim
  \mathcal{N}\!\big(\mathbf{0},\,\Sigma_d\big),
\end{align}
where $W \in \mathbb{R}^{n\times p}$ is the design matrix constructed in Section~\ref{construction_W}, and $\Sigma_d$ denotes the noise covariance of the data. We assume an independent Gaussian noise as $\Sigma_d = \sigma_d^2 I$. 

For the surface map, we place a Gaussian Process (GP) prior to impose spatial smoothness:
\begin{align}
    \boldsymbol{a} \sim
  \mathcal{N}\!\big(\boldsymbol{\mu}_a,\,\Sigma_a\big), \qquad
  \boldsymbol{\mu}_a = \mu_a \mathbf{1},
\end{align}
where $\mu_a$ is a spatially uniform mean brightness prior. 
The covariance matrix $\Sigma_a$ is modeled by a GP kernel based on the angular distance on the sphere.
Specifically, we use the squared-exponential kernel:
\begin{align}
  \Sigma_a[j,j'] = \sigma_a^2 \exp\left(-\frac{d_{jj'}^2}{2\ell^2}\right),
\end{align}
where $d_{jj'}$ is the great-circle distance between pixels $j$ and $j'$, $\sigma_a$ is the amplitude of the surface variations, and $\ell$ is the correlation length scale.

The set of nonlinear parameters determining the system matrices $W$, $\Sigma_d$, and $\Sigma_a$ is collected as
\begin{align}
  \boldsymbol{\theta}
  = (i,\,v_{\mathrm{rot}},\,u,\,\boldsymbol{w},\,\sigma_d,\,\mu_a,\,\sigma_a,\,\ell).
  \label{eq:allparameters}
\end{align}

\subsubsection{Estimating the nonlinear parameters}

Using standard properties of Gaussians under affine transformations and summation, we obtain
\begin{align}
  \mathbb{E}[\boldsymbol{d}]
  &= W\,\mathbb{E}[\boldsymbol{a}]
     + \mathbb{E}[\boldsymbol{\epsilon}]
   = W\,\boldsymbol{\mu}_a, \\[3pt]
  \mathrm{Cov}(\boldsymbol{d})
  &= \mathrm{Cov}(W\boldsymbol{a})
     + \mathrm{Cov}(\boldsymbol{\epsilon})
   = W\,\Sigma_a\,W^\mathsf{T}
     + \Sigma_d.
\end{align}
Hence,
\begin{align}
  p(\boldsymbol{d}\mid\boldsymbol{\theta})
  = \mathcal{N}\!\Bigl(
      \boldsymbol{d}
      \,\Big|\,
      W\,\boldsymbol{\mu}_a,\;
      W\,\Sigma_a\,W^\mathsf{T}
      + \Sigma_d
    \Bigr),
  \label{eq:marginal_likelihood}
\end{align}
so that marginalization over the high-dimensional surface map is available in closed form without explicitly computing any integrals.

The posterior distribution of the nonlinear parameters $\boldsymbol{\theta}$ follows from Bayes’ theorem:
\begin{align}
  p(\boldsymbol{\theta}\mid\boldsymbol{d})
  \propto
  p(\boldsymbol{d}\mid\boldsymbol{\theta})\,p(\boldsymbol{\theta}),
  \label{eq:posterior}
\end{align}
where $p(\boldsymbol{d}\mid\boldsymbol{\theta})$ is the marginal likelihood in Eq.~(\ref{eq:marginal_likelihood}), and $p(\boldsymbol{\theta})$ denotes the prior on the nonlinear parameters.

Because $\boldsymbol{\theta}$ enters the marginal likelihood nonlinearly, the posterior cannot be normalized analytically. We therefore sample $p(\boldsymbol{\theta}\mid\boldsymbol{d})$ using MCMC. With the surface map $\boldsymbol{a}$ analytically marginalized, sampling is restricted to a low-dimensional parameter space and remains efficient even for large $N_{\mathrm{pix}}$. The marginal likelihood retains the full nonlinear dependence on $(i, v_{\mathrm{rot}})$, enabling inference of these parameters directly from the spectra without external constraints. In contrast to classical Doppler imaging, neither $v_{\mathrm{rot}}\sin i$ nor the inclination needs to be fixed a priori.

\subsubsection{Conditional posterior of the surface map}
\label{sec:cond-posterior-a}

Here, $W$, $\Sigma_d$, $\Sigma_a$, and $\boldsymbol{\mu}_a$ are evaluated at a posterior draw $\boldsymbol{\theta}^{(s)}$. With the Gaussian prior
\begin{align}
  \boldsymbol{a}\mid\boldsymbol{\theta} &\sim \mathcal{N}(\boldsymbol{\mu}_a,\Sigma_a),
\end{align}
and the Gaussian likelihood
\begin{align}
  \boldsymbol{d}\mid\boldsymbol{a},\boldsymbol{\theta}
  &\sim \mathcal{N}(W\boldsymbol{a},\Sigma_d),
\end{align}
the conditional distribution $p(\boldsymbol{a}\mid\boldsymbol{d},\boldsymbol{\theta})$ is again Gaussian {(see Appendix \ref{ap:cond_gauss}), expressed as} 
\begin{align}
    p(\boldsymbol{a}\mid\boldsymbol{d},\boldsymbol{\theta})=\mathcal{N}(\boldsymbol{\mu}_{a\mid d},\Sigma_{a\mid d}),
\end{align}
where
\begin{align}
  \Sigma_{a\mid d}
    &= \bigl(\Sigma_a^{-1}+W^\mathsf{T}\Sigma_d^{-1}W\bigr)^{-1},
    \label{eq:posterior-covariance}\\[4pt]
  \boldsymbol{\mu}_{a\mid d}
    &= \Sigma_{a\mid d}\,
       \bigl(W^\mathsf{T}\Sigma_d^{-1}\boldsymbol{d}
             + \Sigma_a^{-1}\boldsymbol{\mu}_a\bigr).
    \label{eq:posterior-mean}
\end{align}

The expression for $\Sigma_{a\mid d}$ in Eq.~(\ref{eq:posterior-covariance}) involves the inversion of the $p\times p$ matrix $\Sigma_a^{-1}+W^\mathsf{T}\Sigma_d^{-1}W$. Whether this inversion is computationally demanding depends on the relative sizes of $p$ (the number of surface elements) and $n$ (the number of data points).

When $p$ is large compared to $n$, it can be advantageous to apply the Woodbury identity to yield the equivalent expressions
\begin{align}
  \Sigma_{a\mid d}
  &= \Sigma_a
     - \Sigma_a W^\mathsf{T}
       \bigl(\Sigma_d + W\Sigma_a W^\mathsf{T}\bigr)^{-1}
       W\Sigma_a,\label{eq:Woodburied-covariance}
  \\
  \boldsymbol{\mu}_{a\mid d}
  &= \boldsymbol{\mu}_a
     + \Sigma_a W^\mathsf{T}
       \bigl(\Sigma_d + W\Sigma_a W^\mathsf{T}\bigr)^{-1}
       (\boldsymbol{d}-W\boldsymbol{\mu}_a).\label{eq:Woodburied-mean}
\end{align}
The detailed derivation is provided in Appendix~\ref{app:woodbury}. These forms require inverting only the $n\times n$ matrix $\Sigma_d + W\Sigma_a W^\mathsf{T}$, and can therefore provide a computational advantage when $n < p$.

\subsubsection{Recovering the marginalized posterior of the surface map}

For each posterior draw $\boldsymbol{\theta}^{(s)}$, the conditional distribution of the surface map, $p(\boldsymbol{a}\mid\boldsymbol{d},\boldsymbol{\theta}^{(s)})$, is Gaussian with mean $\boldsymbol{\mu}_{a\mid d}^{(s)}$ and covariance $\Sigma_{a\mid d}^{(s)}$ given in Eqs.~(\ref{eq:posterior-covariance})--(\ref{eq:posterior-mean}). The marginalized posterior of the surface map is obtained by integrating over the nonlinear parameters:
\begin{align}
  p(\boldsymbol{a}\mid\boldsymbol{d})
  = \int
    p(\boldsymbol{a}\mid\boldsymbol{d},\boldsymbol{\theta})\,
    p(\boldsymbol{\theta}\mid\boldsymbol{d})\,
    d\boldsymbol{\theta}.
  \label{eq:amarg-def}
\end{align}
Since the integral is analytically intractable, we approximate it using the MCMC samples $\{\boldsymbol{\theta}^{(s)}\}_{s=1}^{N_{\mathrm{samples}}}$. Substituting the empirical measure
\(
  p(\boldsymbol{\theta}\mid\boldsymbol{d})
  \approx \frac1{N_{\mathrm{samples}}}
  \sum_{s=1}^{N_{\mathrm{samples}}}
  \delta(\boldsymbol{\theta}-\boldsymbol{\theta}^{(s)})
\)
into Eq.~(\ref{eq:amarg-def}) yields the Monte Carlo approximation
\begin{align}
  p(\boldsymbol{a}\mid\boldsymbol{d})
  \approx
  \frac1{N_{\mathrm{samples}}}
  \sum_{s=1}^{N_{\mathrm{samples}}}
  p(\boldsymbol{a}\mid\boldsymbol{d},\boldsymbol{\theta}^{(s)}),
  \label{eq:amarg-mixture}
\end{align}
i.e., a finite mixture of Gaussians weighted uniformly by the MCMC samples.

While this mixture does not collapse to a single Gaussian, its moments are readily computed.  In particular, the posterior mean and variance of the surface map are approximated by
\begin{align}
  \label{eq:a-mean-mixture}
  \bar{\boldsymbol{\mu}} := \mathbb{E}[\boldsymbol{a}\mid\boldsymbol{d}]
  &\approx
  \frac1{N_{\mathrm{samples}}}
  \sum_{s=1}^{N_{\mathrm{samples}}}
  \boldsymbol{\mu}_{a\mid d}^{(s)}, \\
\label{eq:a-variance-mixture}
  \mathrm{Var}[\boldsymbol{a}\mid\boldsymbol{d}] 
  &\approx
    \frac1{N_{\mathrm{samples}}} \nonumber \\
  &\times \sum_{s=1}^{N_{\mathrm{samples}}}
    \bigl[
      \Sigma_{a\mid d}^{(s)}
      +
      (\boldsymbol{\mu}_{a\mid d}^{(s)}-\bar{\boldsymbol{\mu}})
      (\boldsymbol{\mu}_{a\mid d}^{(s)}-\bar{\boldsymbol{\mu}})^\mathsf{T}
    \bigr].
\end{align}

\section{Synthetic Data Experiments}
\label{sec:synthetic}
To test the Bayesian Doppler imaging framework, we generate synthetic phase-resolved spectra from representative surface-brightness maps using the forward model described in Section~\ref{discretizing}.

\subsection{Ground Truth Maps and Mock Spectra} 
\label{sec:mock}
The stellar surface is discretized on a HEALPix grid \citep{2005ApJ...622..759G} with $N_{\mathrm{side}}=8$ ($N_{\mathrm{pix}}=768$). We construct three brightness maps with varying spatial scales, asymmetries, and numbers of features. Each map consists of circular spots, defined by $(\theta,\phi)$, angular radius, and attenuation, superposed on a uniform surface.

\paragraph{Map~1 (single mid-latitude spot)}
A single circular spot of radius $\pi/6$ is placed at colatitude $\theta=\pi/4$ and longitude $\phi=0$, providing a minimal test case with a localized feature.

\paragraph{Map~2 (two-spot configuration)}
A spot of radius $\pi/6$ at $\theta=\pi/3$, $\phi=\pi/4$ is combined with a larger spot of radius $\pi/5$ on the opposite hemisphere ($\theta=2\pi/3$, $\phi=-2\pi/3$), introducing longitudinal asymmetry and multiple spatial scales.

\paragraph{Map~3 (four irregularly distributed spots)}
Map~2 is extended with two smaller spots of radius $\pi/8$ at $(\theta=\pi/6,\phi=-3\pi/4)$ and $(\theta=7\pi/12,\phi=3\pi/4)$, forming a complex, non-axisymmetric surface pattern.

The spectra are evaluated on a uniform wavelength grid of 100 points covering $\lambda_0\pm 0.15\,\mathrm{nm}$ around $\lambda_0=656.28\,\mathrm{nm}$.  The intrinsic line profile is a Gaussian absorption line of depth 0.8 in wavenumber space:
\begin{align}
  s^\ast(\lambda)
 &= 1 - 0.8\,\exp\!\left[-\tfrac12(\nu-\nu_0)^2/\sigma^2\right],\\
  \nu &= \frac{10^7}{\lambda},\ \sigma = 0.3.
\end{align}
This provides a smooth, symmetric line shape suitable for controlled experiments.

We generate spectra at $N_{\mathrm{phase}}=8$ equally spaced rotational phases.  To emulate realistic variations in exposure or instrumental throughput, each spectrum is scaled by a small phase-dependent weight, $
%\begin{align}
  \boldsymbol{w} = (1.00,\,0.98,\,1.03,\,0.99,\,1.01,\,0.97,\,1.02,\,1.00)
%\end{align}
$, corresponding to the eight phases.

To specifically assess whether the degeneracy between inclination $i$ and the equatorial rotation velocity $v_{\mathrm{rot}}$ can be disentangled using the spectra, we fix the projected rotation velocity $v_{\mathrm{rot}}\sin i$ to $10~\mathrm{km\,s^{-1}}$ and vary the inclination from $10^\circ$ to $80^\circ$ in steps of $10^\circ$. The corresponding equatorial rotation speeds are set to
\begin{align}
  v_{\mathrm{rot,true}}
  = \frac{10 \, \mathrm{km\,s^{-1}}}{\sin i_{\mathrm{true}}}.
\end{align}
The limb-darkening coefficient is fixed at $u=0.5$ for all synthetic realizations.

For each parameter pair $(i_{\mathrm{true}},v_{\mathrm{rot,true}})$ and each map, a noiseless spectrum is generated and subsequently perturbed with additive Gaussian noise:
\begin{align}    
  \boldsymbol{d} = \boldsymbol{d}_{\mathrm{true}} + \boldsymbol{\epsilon},
  \qquad
  \boldsymbol{\epsilon}\sim\mathcal{N}(\boldsymbol{0},\sigma^2 I).
\end{align}
The noise variance is set to 
\begin{align}  
  \sigma = 0.02\,\lVert \boldsymbol{d}_{\mathrm{true}} \rVert_\infty,
\end{align}
corresponding to a 2\% relative noise level. This ensures that the data quality is representative of high-S/N spectroscopic observations while still posing a nontrivial inference challenge.

The resulting noisy spectra, together with the fixed intrinsic line profile, serve as the inputs to the Bayesian inference described in Section~\ref{sec:inference-setup}.

As defined in Eq.~(\ref{eq:allparameters}), the nonlinear parameters are
\begin{align}
  \boldsymbol{\theta}
  = (i,\,v_{\mathrm{rot}},\,u,\,\boldsymbol{w},\,\sigma_d,\,\mu_a,\,\sigma_a,\,\ell).
\end{align}
which determine the forward operator $W(\boldsymbol{\theta})$, the noise covariance $\Sigma_d(\boldsymbol{\theta})$, and the prior covariance of the surface map $\Sigma_a(\boldsymbol{\theta})$. The priors for each parameter are provided in Table \ref{tab:priors}.

\subsection{Inference settings}
\label{sec:inference-setup}
\begin{table}[htbp]
\centering
\caption{Priors used for the nonlinear parameters in the synthetic data experiments. For the inclination, we impose an isotropic prior on the spin axis, which corresponds to a uniform prior on $\cos i$. The parameters $w_k$ and $\ell$ are modeled through their logarithms in order to ensure positivity.}
\begin{tabular}{lll}
\hline\hline
 \multicolumn{2}{l}{Symbol and meaning} & Prior \\
\hline
$i$              & Inclination (isotropic prior) &
  $\cos i\sim\mathrm{Uniform}(0,1)$ \\
$v_{\mathrm{rot}}$    & Equatorial rotation velocity &
  $\mathrm{Uniform}(0,60)$ \\
$u$                   & Limb-darkening coefficient &
  $\mathrm{Uniform}(0,1)$ \\
$w_k$            & Phase-dependent weights&
  $\log w_k\sim\mathcal{N}(0,0.1^2)$ \\[3pt]
$\sigma_d$            & Noise amplitude &
  $\mathrm{HalfNormal}(10)$ \\[3pt]
$\mu_a$               & Mean surface brightness &
  $\mathrm{Beta}(2,2)$ \\
$\sigma_a$            & GP amplitude; $\Sigma_a \propto \sigma_a^2$ &
  $\mathrm{HalfNormal}(0.3)$ \\
$\ell$                & GP correlation length scale &
  $\log \ell \sim \mathcal{N}(-1,0.5^2)$ \\
\hline
\end{tabular}
\label{tab:priors}
\end{table}

Posterior sampling is conducted using the Hamiltonian Monte Carlo and No-U-Turn Sampler \citep[HMC-NUTS; e.g.][]{1987PhLB..195..216D,2011arXiv1111.4246H,2017arXiv170102434B} in {\sf NumPyro} \citep{2019arXiv191211554P} with 500 warm-up iterations and 1000 posterior samples, employing a dense mass matrix and a target acceptance probability of~0.9.  At each MCMC step, the design matrix $W(\boldsymbol{\theta})$ is recomputed from the current values of $(i, v_{\mathrm{rot}}, u, \boldsymbol{w})$.

\subsection{Recovery of the nonlinear parameters}
\label{sec:recovery-params}

\begin{figure*}[htbp]
  \centering
  \includegraphics[width=\linewidth]{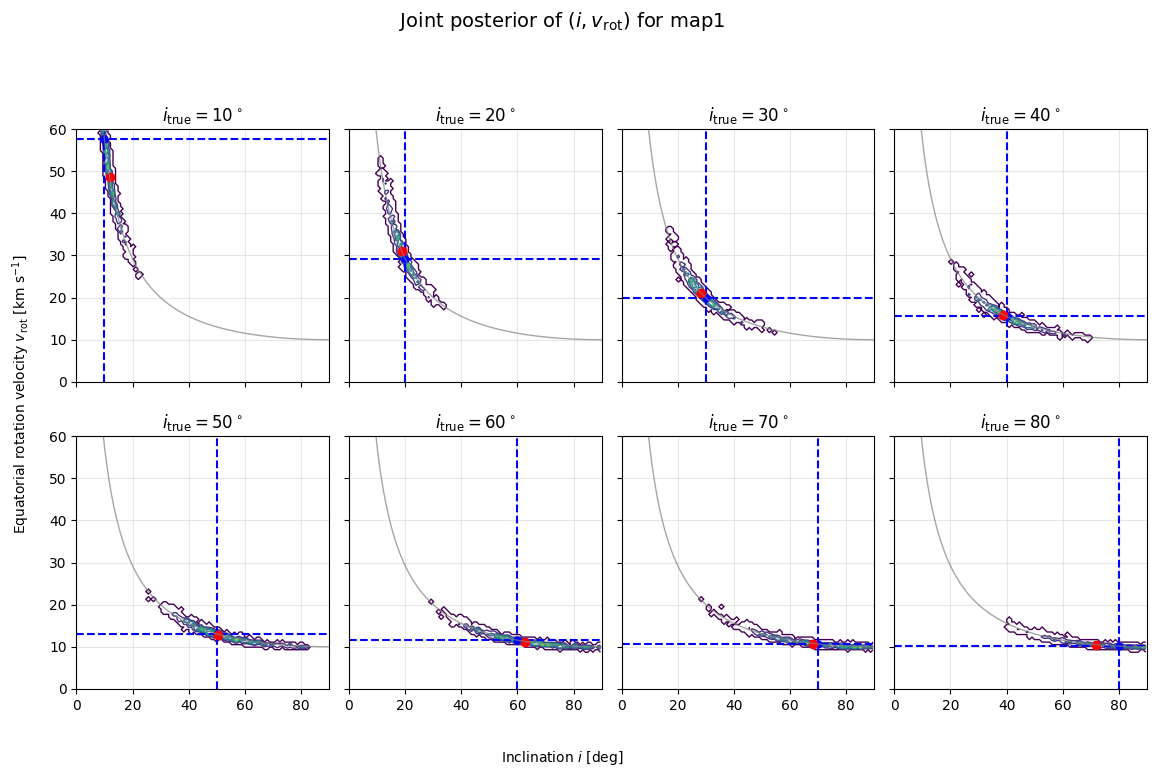}
  \caption{Joint posterior distributions of $(i, v_{\mathrm{rot}})$ inferred from synthetic spectra for each of the true inclinations $i = 10^\circ, 20^\circ, \ldots, 80^\circ$ using Map~1. Blue dashed lines indicate the injected (true) values, red points show the posterior medians, and thin gray curves denote the locus $v_{\mathrm{rot}}\sin i = 10~\mathrm{km\,s^{-1}}$.}
  \label{fig_i-v_map=map1}
\end{figure*}

We examine the recovery of the geometric parameters $(i, v_{\mathrm{rot}})$ from the synthetic spectra.
Figure~\ref{fig_i-v_map=map1} shows the joint posterior distributions for Map~1 at all injected inclinations. The posteriors are elongated along $v_{\mathrm{rot}}\sin i \approx 10~\mathrm{km,s^{-1}}$, reflecting the strong constraint from the projected rotation velocity. However, the samples cluster tightly around the true parameter values $(i_{\mathrm{true}}, v_{\mathrm{rot,true}})$ in each case. Thus, despite a residual degeneracy in shape, both parameters are accurately recovered from the phase-resolved spectra.

\begin{figure*}[htbp]
  \centering
  \includegraphics[width=\linewidth]{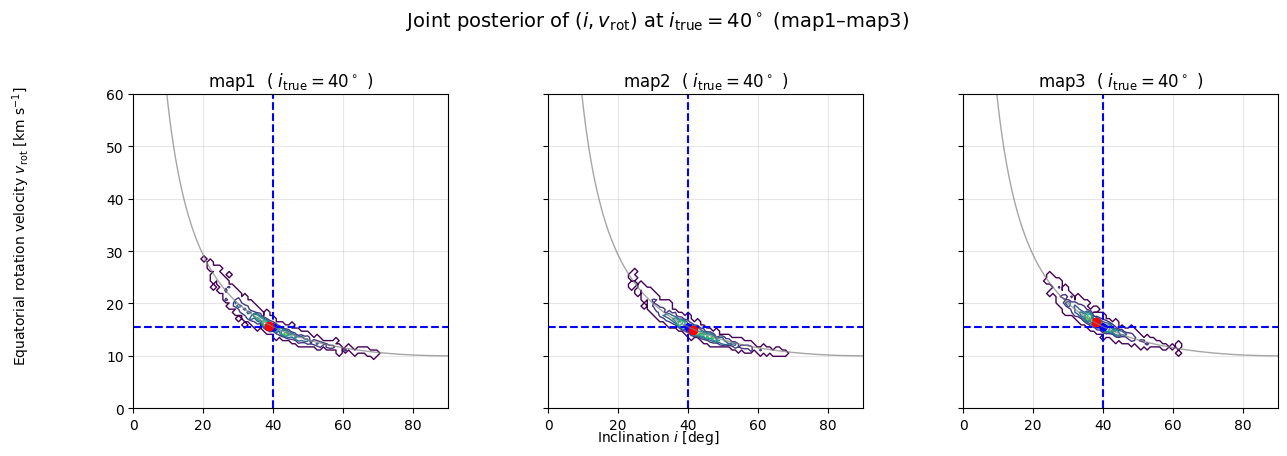}
  \caption{Joint posterior distributions of $(i, v_{\mathrm{rot}})$ inferred from synthetic spectra at a fixed inclination of $i = 40^\circ$ for Maps~1, 2, and~3. Blue dashed lines indicate the injected (true) values, red points show the posterior medians, and thin gray curves denote the locus $v_{\mathrm{rot}}\sin i = 10~\mathrm{km\,s^{-1}}$.}
  \label{fig_i-v_i=40deg}
\end{figure*}

Figure~\ref{fig_i-v_i=40deg} compares the posteriors for Maps~1–3 at a fixed inclination of $i=40^\circ$. The constraints tighten with increasing surface complexity, with Map~3 yielding the narrowest posteriors. More structured surfaces produce stronger phase-dependent modulation, further reducing the residual $v$–$i$ degeneracy.
The hyperparameters are also well constrained: the noise amplitude $\sigma_d$ recovers the injected 2\% level, and the GP correlation length $\ell$ adapts to the characteristic spot size of each map.

\subsection{Recovery of the surface map}
\label{sec:recovery-map}
\begin{figure*}[htbp]
  \centering
  \includegraphics[width=\linewidth]{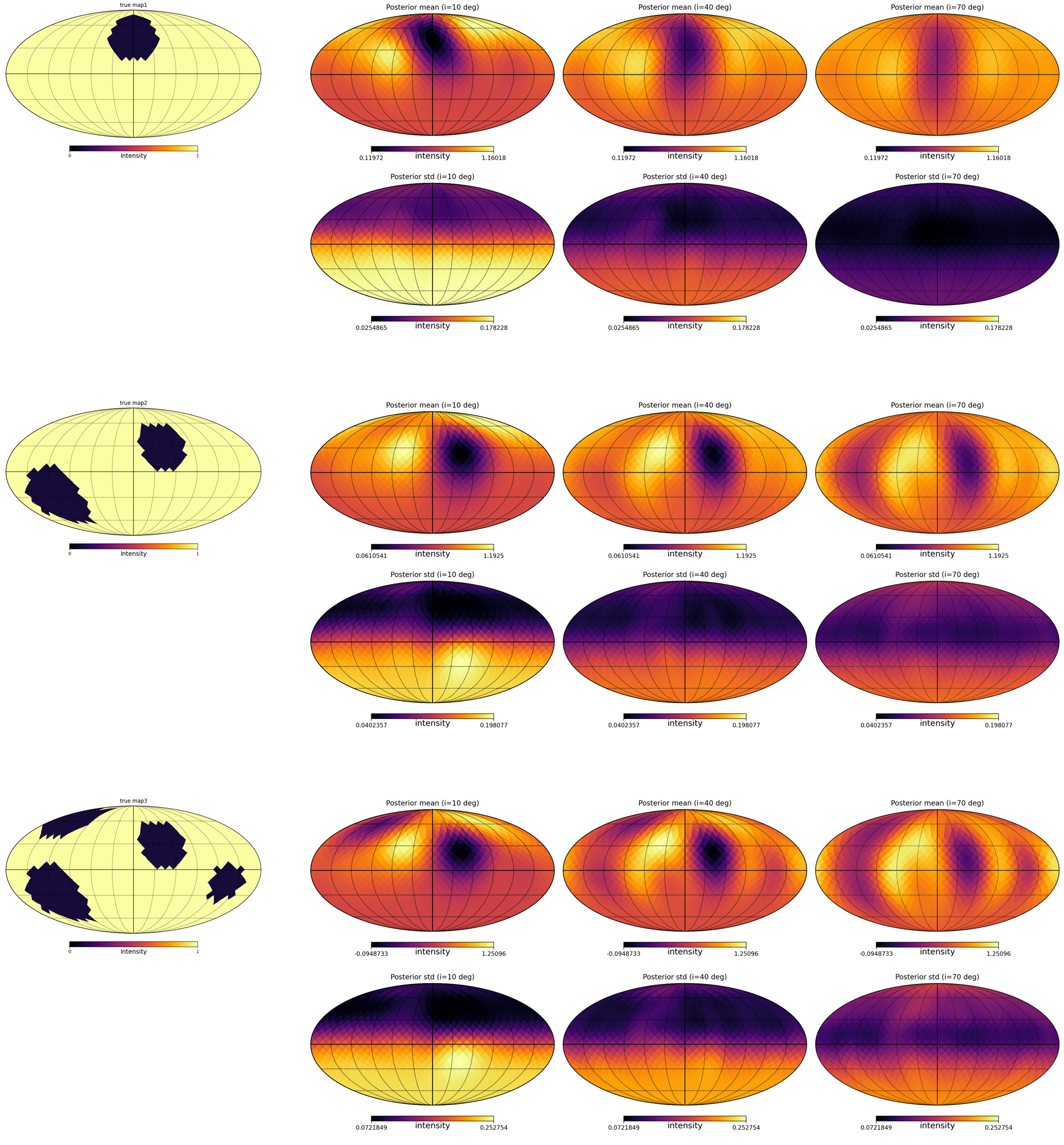}
  \caption{Posterior mean (top) and uncertainty (bottom) surface maps inferred from synthetic spectra. The rows show the results for Map~1, Map~2, and Map~3, respectively. The first column presents the ground truth, while the remaining columns display the reconstructions at inclinations of $i=10^{\circ}$, $40^{\circ}$, and $70^{\circ}$, shown on a common color scale within each row.}
  \label{fig:posterior_means&std}
\end{figure*}

Figure~\ref{fig:posterior_means&std} shows the posterior mean maps (Eq.~\ref{eq:a-mean-mixture}) for Maps~1--3 at inclinations $i=10^\circ$, $40^\circ$, and $70^\circ$.

\paragraph{Map~1 (single mid-latitude spot)}
The spot is recovered at all inclinations. Its longitude is well constrained, while the reconstruction is elongated in latitude. At low inclinations, the apparent agreement in latitude reflects the limited visibility of southern regions rather than true latitudinal resolution, confirming that latitude is intrinsically weakly constrained. {In addition, spurious bright spots are introduced surrounding the dark spot. This could be related to the behavior of the Gaussian Process prior, as the requirement for smoothness can sometimes induce a subtle overshoot when modeling sharp gradients.}

\paragraph{Map~2 (two spots)}
At low inclinations, only the northern spot leaves a detectable signal and is recovered, while the southern spot is largely missed. As $i$ increases, both spots become visible and are reconstructed; { however, the southern spot is only faintly recovered compared to the northern one}. Latitudinal elongation persists, indicating stronger sensitivity to longitudinal structure than to latitude, {and the recovery of southern 
features remains limited when they are viewed only weakly.}

\paragraph{Map~3 (four spots)}
Increasing surface complexity does not { significantly} degrade reconstruction quality, {though it still remains highly dependent on inclination. At $i=10^\circ$, only the northern spots are recovered, while the southern features remain hidden.} At intermediate inclinations ($i \approx 40^\circ$--$60^\circ$), spots at similar longitudes can be partially separated, showing that favorable viewing geometries provide sufficient Doppler leverage to distinguish overlapping features. {However, at $i=70^\circ$, this separation is lost as latitudinal elongation causes the features to blend together. In both cases, the southern features is only faintly recovered compared to the northern one.}

Figure~\ref{fig:posterior_means&std} also shows the spatial uncertainties, given by the square roots of the diagonal elements of the covariance matrices (Eq.~\ref{eq:a-variance-mixture}). The uncertainties are smallest near the visible pole at low inclinations and shift toward lower latitudes as $i$ increases, reflecting the limited visibility of the opposite hemisphere at low $i$. {Notably, in the cases of Map 2 and Map 3 at \(i = 10^\circ\), 
enhanced uncertainties are seen in the southern hemisphere, including 
regions near the same longitudes as the recovered northern spots. This 
does not imply the detection of mirror features. Instead, the enhanced 
uncertainty mainly reflects the low-inclination viewing geometry: the 
southern hemisphere is poorly visible, contributes weakly to the 
line-profile variations, and therefore remains only weakly constrained 
by the data. The same-longitude appearance of the uncertainty should 
therefore be regarded as a manifestation of limited visibility and weak 
latitudinal leverage rather than as a physical counterpart to the 
northern spots. Even in the least constrained regions, the standard 
deviation remains below 20\% of the posterior mean.}

{These tests show that the method is most reliable for identifying the longitudes of large, dark surface inhomogeneities. In contrast, the detailed latitudinal structure and isolated bright patches should be interpreted with caution. In particular, bright regions appearing adjacent to or opposite large dark features can arise as reconstruction artifacts associated with the limited latitudinal sensitivity of Doppler imaging and the smooth GP prior.}

\section{Application to Luhman 16B}
\label{sec:application}

We apply our method to actual high-dispersion spectroscopic observations of the brown dwarf Luhman 16B, which was the first brown dwarf to which Doppler imaging was applied by \citet{2014Natur.505..654C}. The data were acquired on May 5, 2013, using the CRIRES (Cryogenic High-Resolution Infrared Echelle Spectrograph) instrument mounted on the ESO Very Large Telescope (VLT).

\subsection{Observations and Data reduction}
\label{subsec:obs_and_reduction}
We analyze the spectral data used in \citet{2014Natur.505..654C}. The data cover the $K$-band wavelength range of $2.28$--$2.35\,\mu\mathrm{m}$ across four detector chips. In this work, we specifically focused on the data recorded on Chip 2 ($2.304$--$2.316\,\mu\mathrm{m}$), which contains prominent molecular absorption features of the CO bands. The observations were carried out with a slit width of 0.2 arcseconds, yielding a spectral resolution of $R \approx 100,000$. The target was monitored continuously for approximately 5 hours, corresponding to one full rotation period of Luhman 16B. The resulting time series consists of 14 spectra ($N_\mathrm{phase}=14$), which capture the spectral variability induced by its patchy cloud cover.

\subsection{Model setup}
\label{subsec:model_setup}

The surface was discretized using a HEALPix grid with $N_{\mathrm{side}}=8$, resulting in $N_{\mathrm{pix}}=768$ pixels. 
To generate the intrinsic line profiles, we adopt the atmospheric model from the retrieval analysis of \citet{2025arXiv251123018Y}, who analyzed the same VLT/CRIRES observations of Luhman~16B using the differentiable spectral modeling code {\sf ExoJAX} \citep{Kawahara_2022, Kawahara_2025}. We compute the posterior predictive spectrum with the projected rotational velocity fixed to $v_{\mathrm{rot}}\sin i = 0~\mathrm{km\,s^{-1}}$, based on the MCMC posterior samples. The median of this predictive distribution is adopted as the intrinsic line profile ($\boldsymbol{s}^\ast$ in Section~\ref{construction_W}), yielding a thermally and pressure-broadened rest-frame spectrum. This spectrum is used as the local intensity kernel for the Doppler imaging inversion.

The Bayesian inference was performed for a set of parameters similar to that in the synthetic data experiments (Section~\ref{sec:inference-setup}), but with modifications to the limb-darkening model and the specific prior distributions to accommodate the real data properties. Unlike the synthetic experiments where the rotational phases were fixed, the phases for Luhman 16B depend on the rotation period $P$, which involves uncertainty. Therefore, we explicitly include the rotation period $P$ in the set of inferred parameters to calculate the rotational phases from the observation timestamps. Additionally, we adopted a quadratic limb-darkening law for Luhman 16B to maintain consistency with the intrinsic line profile generated by the atmospheric model (although we note that preliminary runs using a linear limb-darkening law yielded virtually identical results).
Consequently, the set of inferred nonlinear parameters is
\begin{align}
  \boldsymbol{\theta} = (i,\,v_{\mathrm{rot}},\,q_1,\, q_2,\,\boldsymbol{w},\,\sigma_d,\,\mu_a,\,\sigma_a,\,\ell, P),
\end{align}
where $q_1$ and $q_2$ are the transformed quadratic limb-darkening coefficients \citep{2013MNRAS.435.2152K}, and $\boldsymbol{w}$ is a vector of length 14 corresponding to the number of observed spectra ($N_{\mathrm{phase}}=14$). 

Given the differences in data quality and physical context compared to the synthetic experiments, we adjusted several prior distributions. The priors used for the application to Luhman 16B are summarized in Table~\ref{tab:priors_luhman16b}.

\begin{table}[htbp]
\centering
\caption{Priors used for the application to Luhman 16B.}
\begin{tabular}{lll}
\hline\hline
Parameter & Prior \\
\hline
$i$ & $\cos i\sim\mathrm{Uniform}(0,1)$ \\
$v_{\mathrm{rot}}$ & $\mathrm{Uniform}(0,120)$ \\
$q_1, q_2$ & $\mathrm{Uniform}(0,1)$ \\
$w_k$ & $\log w_k\sim\mathcal{N}(0,0.1^2)$ \\
$\sigma_d$ & $\mathrm{LogNormal}(\ln 0.03, 1.0^2)$ \\
$\mu_a$ & $\mathrm{Uniform}(0, 0.05)$ \\
$\sigma_a$ & $\mathrm{HalfNormal}(0.3)$ \\
$\ell$ & $\mathrm{Uniform}(0.1, 1.5)$ \\
$P$ & $\mathrm{Uniform}(4.8, 5.4)$ \\
\hline
\end{tabular}
\label{tab:priors_luhman16b}
\end{table}

\subsection{Inference of the nonlinear parameters}
\label{sec:recovery-params2}

\begin{figure}[htbp]
  \centering
  \includegraphics[width=\linewidth]{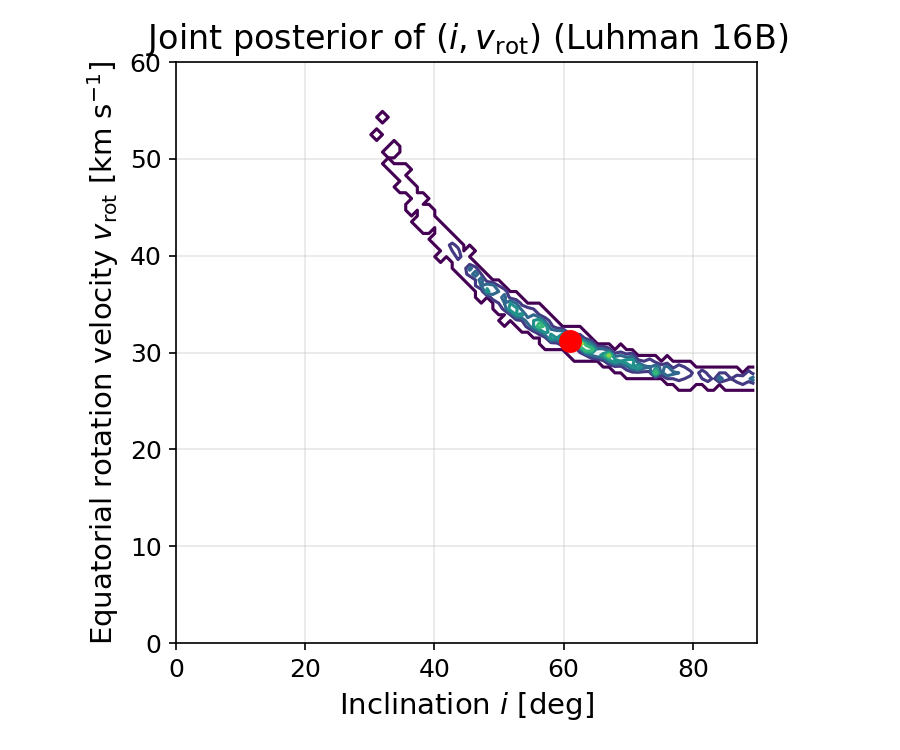}
  \caption{Joint posterior distributions of $(i, v_{\mathrm{rot}})$ inferred from Luhman 16B data. The red point shows the posterior median.}
  \label{fig_i-v_luh16B}
\end{figure}

Figure~\ref{fig_i-v_luh16B} shows the joint posterior distribution of $v_{\mathrm{rot}}$ and $i$, which exhibits the strongest correlation among the geometric parameters. Despite the conservative prior adopted for the real-data application, both parameters are well constrained by the time-resolved spectral variations. The equatorial rotation velocity is inferred to be $v_{\mathrm{rot}} = 31.2_{-3.1}^{+5.3}~\mathrm{km\,s^{-1}}$, while the inclination is constrained to $i = {61.0}_{-12.3}^{+14.3}$ degrees. These estimates correspond to a projected rotational velocity of $v_{\mathrm{rot}} \sin i \approx 27.3\,\mathrm{km\,s^{-1}}$.

\begin{figure}[htbp]
  \centering
  \includegraphics[width=\linewidth]{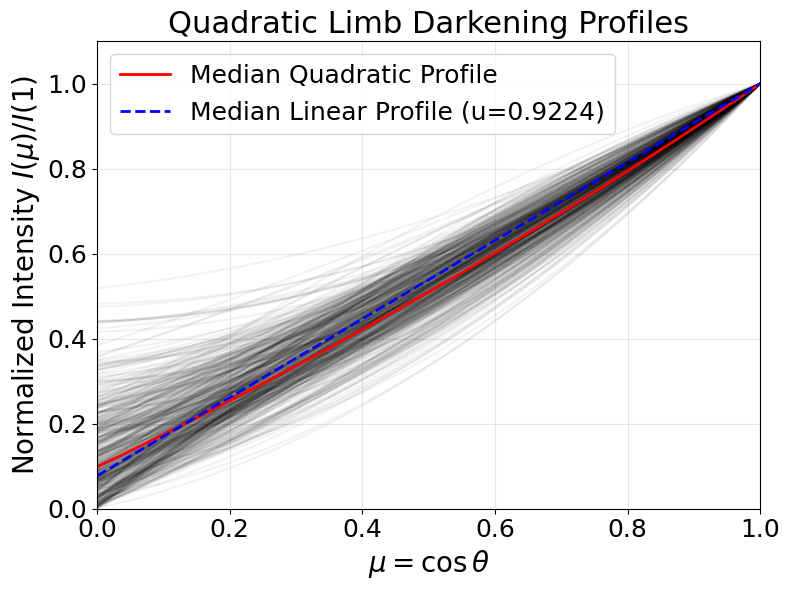}
  \caption{Reconstructed limb-darkening profiles derived from the posterior samples of the quadratic coefficients $(q_1, q_2)$. The red line represents the median, while the thin gray lines show 1000 individual profiles drawn from the posterior distribution, illustrating the uncertainty. For comparison, the blue dashed line shows the median profile obtained from a separate preliminary run assuming a linear limb-darkening law ($u \approx 0.92$).}
  \label{fig:ld_profile}
\end{figure}

We also examine the limb darkening coefficients $q_1$ and $q_2$, yielding estimates of $q_1 = 0.81_{-0.22}^{+0.13}$ and $q_2 = 0.59_{-0.13}^{+0.16}$. Figure \ref{fig:ld_profile} displays the limb-darkening profiles reconstructed from the posterior samples of $q_1$ and $q_2$. The bundle of profiles (shown in gray) with their median (red line) demonstrates that the center-to-limb intensity variation is tightly constrained by the data. To assess the sensitivity of our results to the choice of the limb-darkening law, we compare this result with the median profile obtained from a linear limb-darkening model (blue dashed line), which yields a coefficient of $u \approx 0.92$. The excellent agreement between the quadratic and linear profiles indicates that the inferred surface map and geometric parameters are robust against the specific parameterization of the limb darkening.

The noise amplitude is inferred to be $\sigma_d = 0.0387_{-0.0003}^{+0.0002}$, with a sharply peaked posterior. This indicates that the overall noise level is tightly constrained by the data within the assumed homoscedastic Gaussian noise model.

\begin{figure*}[htbp]
  \centering
  \includegraphics[width=\linewidth]{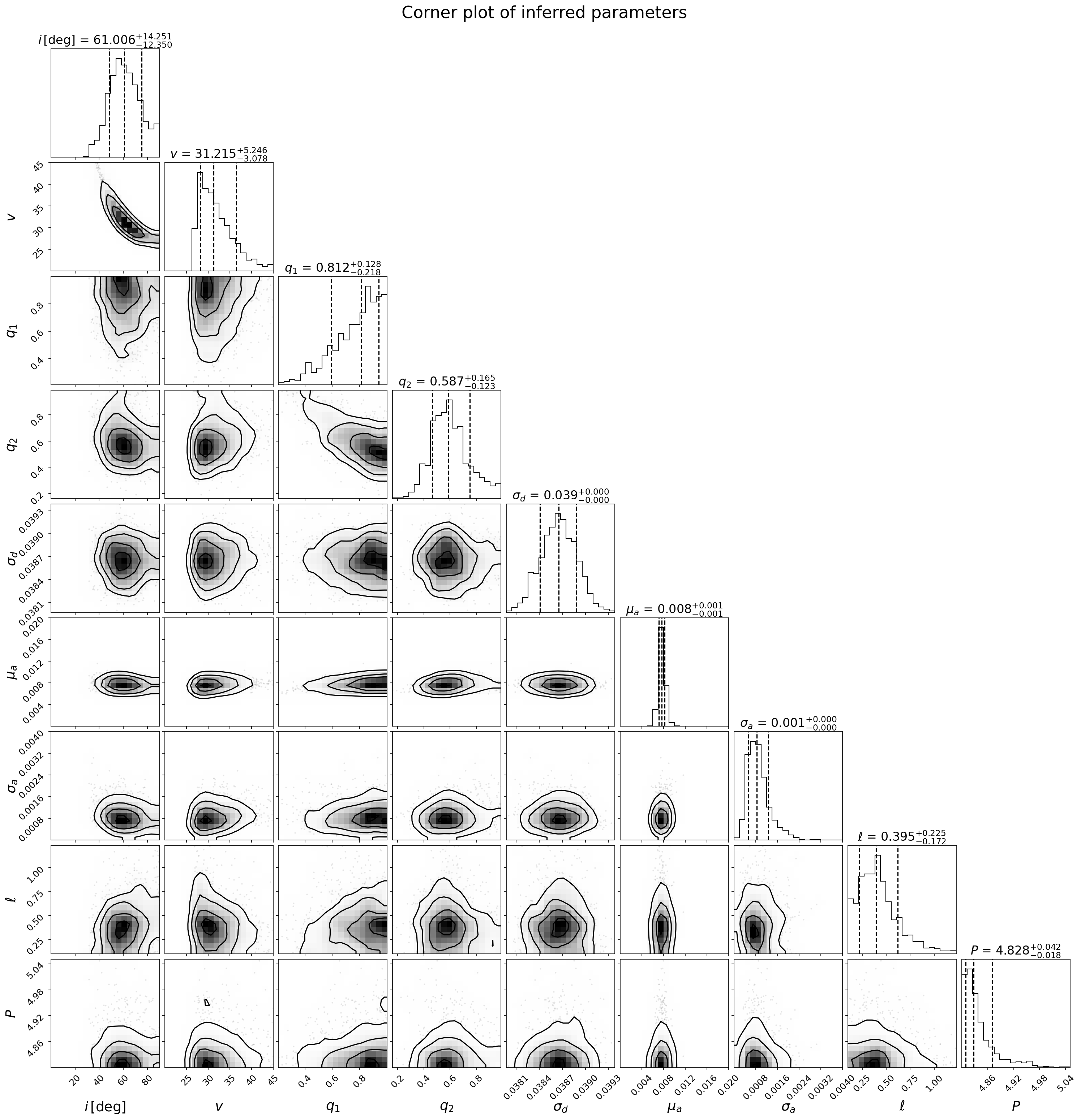}
  \caption{Corner plot showing the one- and two-dimensional posterior distributions for all nonlinear parameters, excluding the regularization weight $w_{k}$. The diagonal panels display the marginalized probability density for each parameter, while the off-diagonal panels show the joint distributions. The vertical lines in the diagonal panels indicate the 16th, 50th, and 84th percentiles.}
  \label{fig_corner_all}
\end{figure*}

Figure~\ref{fig_corner_all} summarizes the posterior distributions for the full set of nonlinear parameters, excluding the regularization weight $w$. Apart from the correlations discussed above, this global view confirms that the remaining parameter pairs do not exhibit significant degeneracies, ensuring the robustness of the solution.
%A detailed summary of the posterior estimates for all scalar parameters is provided in Appendix~\ref{app:allparams_luh16B}.

Overall, these results demonstrate that the nonlinear parameters governing the Doppler imaging forward model are jointly constrained by the Luhman~16B data. This establishes a well-defined set of model parameters that underpins the surface reconstruction presented in the following section.

\subsection{Recovery of the surface map}
\label{sec:recovery-map2}
\begin{figure}[htbp]
  \centering
  \includegraphics[width=\linewidth]{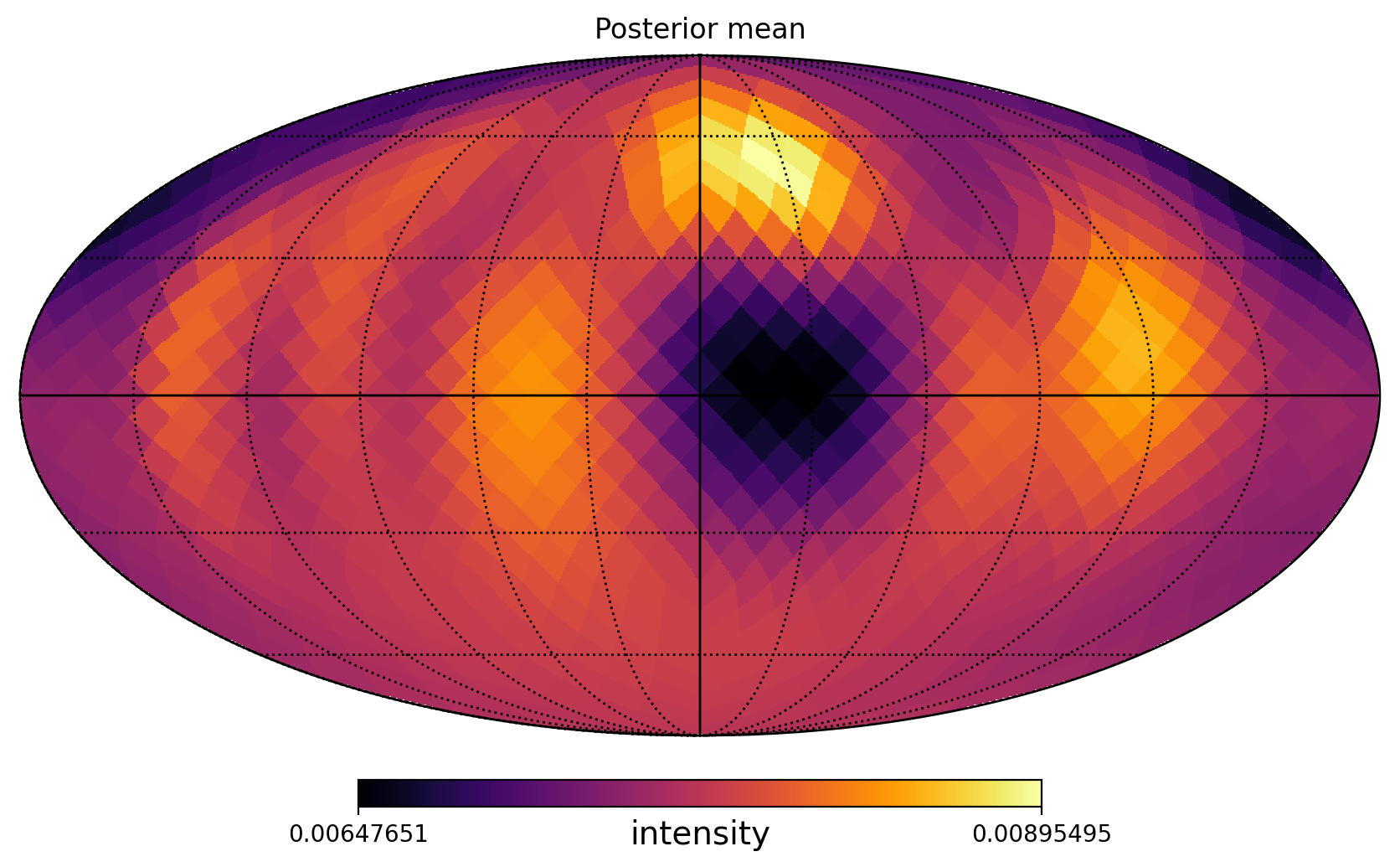}
  \includegraphics[width=\linewidth]{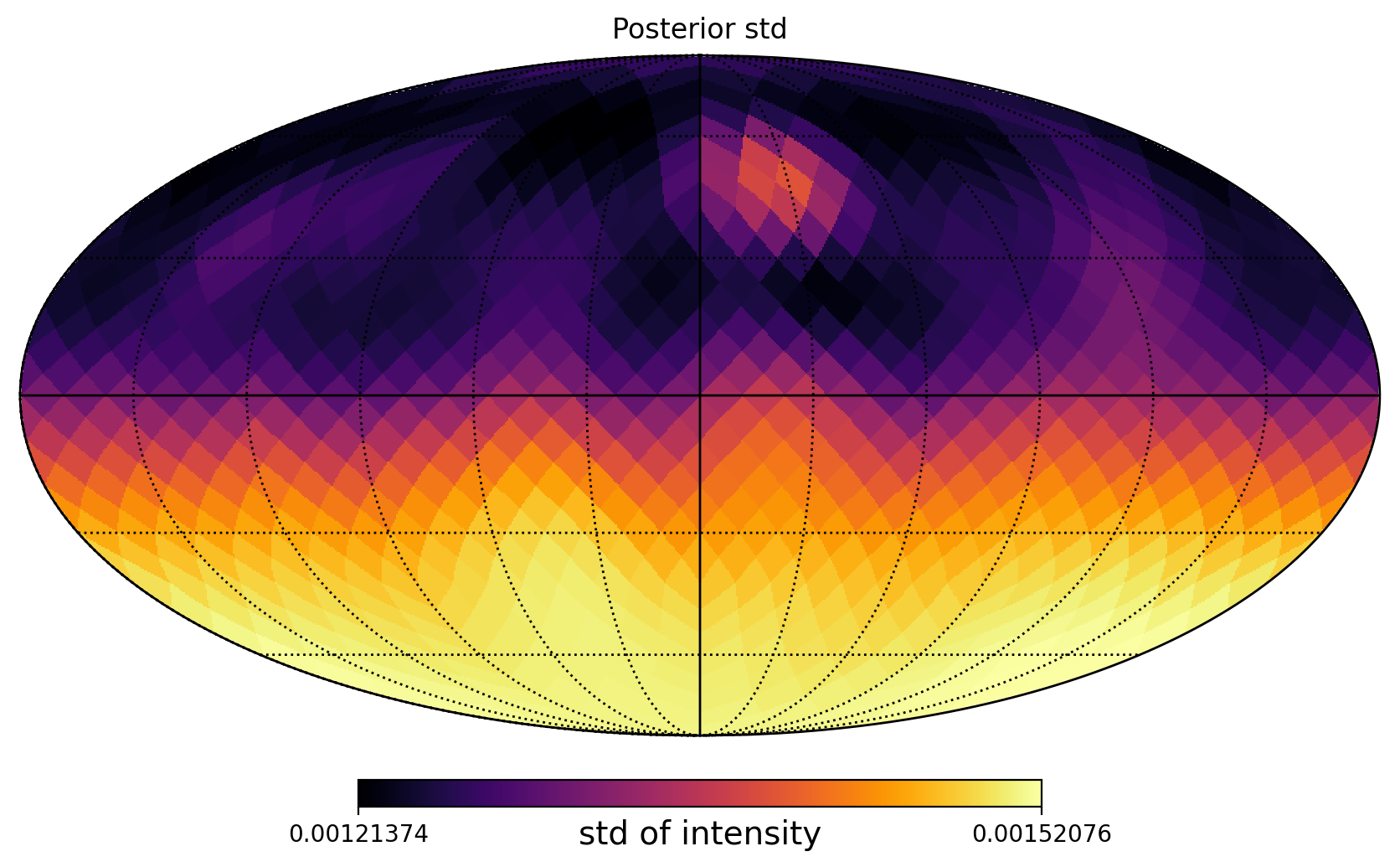}
  \caption{Posterior mean (top) and uncertainty (bottom) maps for Luhman 16B.}
    \label{fig:map_luh16B}
\end{figure}

Using the posterior samples of the nonlinear parameters, we reconstruct the surface brightness distribution of Luhman~16B. Figure~\ref{fig:map_luh16B} (top) shows the posterior mean surface map and Figure~\ref{fig:map_luh16B} (bottom) shows the corresponding pixel-wise posterior standard deviation.

The posterior mean map reveals significant inhomogeneities on the surface of Luhman 16B. The most prominent feature is a large dark region with substantial latitudinal extent located at a longitude of approximately $\phi^\ast=\pi$. Note that this feature appears split across the left and right edges of the Mollweide map due to the projection{; instead, Figure~\ref{fig:map_comparison} (top) offers a continuous view of this region with a 180° longitudinal shift}. Additionally, we identify a spot-like feature near the equator at a longitude of roughly $\phi^\ast=\pi/6$, which is accompanied by a bright region of comparable size located at a slightly higher latitude. {However, as demonstrated in our synthetic data experiments, this bright region may be a spurious feature rather than a physical structure. The relative brightness contrast of these structures reaches approximately 10\% relative to the mean surface level. }

{Guided by the synthetic experiments, we regard the large dark region 
near \(\phi^\ast \simeq \pi\) as the most robust feature of the 
reconstruction. Its large spatial extent and relatively low posterior 
standard deviation suggest that this feature is driven by the data 
rather than by pixel-scale noise. By contrast, the bright region near 
\(\phi^\ast \simeq \pi/6\), as well as isolated high-latitude bright 
patches, should be interpreted with caution. Similar bright patches 
appear in the synthetic reconstructions even when no bright spots are 
injected, and may therefore arise from latitudinal degeneracy and the 
smooth GP prior.}

Consistent with our synthetic data experiments, the uncertainty is relatively small across the hemisphere containing the visible pole, where the observational coverage is most dense. In particular, we find that the standard deviation is notably low in the vicinity of the large dark region at longitude $\phi^\ast\sim \pi$. This indicates that the spectral signal from this prominent feature is well-constrained by the data, further reinforcing the reliability of its detection.

Overall, the recovered surface map and its associated uncertainty demonstrate that the Bayesian Doppler imaging framework yields a well-defined posterior distribution over surface brightness even when applied to real spectroscopic data. The presence of coherent structures in the posterior mean, together with spatially structured uncertainty, indicates that the inferred map is driven by the data rather than by the prior alone.

\subsection{Spectral fit and residuals}
\label{sec:fit}

\begin{table}[htbp]
  \centering
  \caption{Posterior estimates of the nonlinear parameters for Luhman 16B.}
  \label{tab:luhman16b_params}
  \begin{tabular}{lcccc}
    \hline
    Parameter & Mean & Median & 16\% & 84\% \\
    \hline
    $\cos i$ & 0.4630 & 0.4847 & 0.2545 & 0.6606 \\
    $v_{\mathrm{rot}}$ & 32.3945 & 31.2152 & 28.1371 & 36.4613 \\
    $q_1$ & 0.7734 & 0.8124 & 0.5947 & 0.9403 \\
    $q_2$ & 0.6012 & 0.5873 & 0.4641 & 0.7525 \\
    $\log w_1$ & -0.0102 & -0.0101 & -0.0374 & 0.0187 \\
    $\log w_2$ & -0.0231 & -0.0223 & -0.0507 & 0.0054 \\
    $\log w_3$ & -0.0222 & -0.0204 & -0.0489 & 0.0060 \\
    $\log w_4$ & -0.0129 & -0.0116 & -0.0399 & 0.0158 \\
    $\log w_5$ & 0.0023 & 0.0029 & -0.0256 & 0.0305 \\
    $\log w_6$ & 0.0155 & 0.0162 & -0.0127 & 0.0427 \\
    $\log w_7$ & 0.0194 & 0.0192 & -0.0090 & 0.0469 \\
    $\log w_8$ & 0.0078 & 0.0083 & -0.0198 & 0.0343 \\
    $\log w_9$ & -0.0094 & -0.0084 & -0.0359 & 0.0177 \\
    $\log w_{10}$ & -0.0141 & -0.0129 & -0.0404 & 0.0140 \\
    $\log w_{11}$ & -0.0067 & -0.0062 & -0.0343 & 0.0214 \\
    $\log w_{12}$ & 0.0026 & 0.0027 & -0.0246 & 0.0318 \\
    $\log w_{13}$ & -0.0022 & -0.0027 & -0.0300 & 0.0275 \\
    $\log w_{14}$ & -0.0148 & -0.0138 & -0.0420 & 0.0142 \\
    $\sigma_d$ & 0.0387 & 0.0387 & 0.0384 & 0.0389 \\
    $\mu_a$ & 0.0077 & 0.0077 & 0.0071 & 0.0082 \\
    $\sigma_a$ & 0.0009 & 0.0008 & 0.0005 & 0.0013 \\
    $\ell$ & 0.4237 & 0.3945 & 0.2227 & 0.6196 \\
    $P$ & 4.8404 & 4.8278 & 4.8094 & 4.8699 \\
    \hline
  \end{tabular}
\end{table}

\begin{figure*}[htbp]
  \centering
  \includegraphics[width=\linewidth]{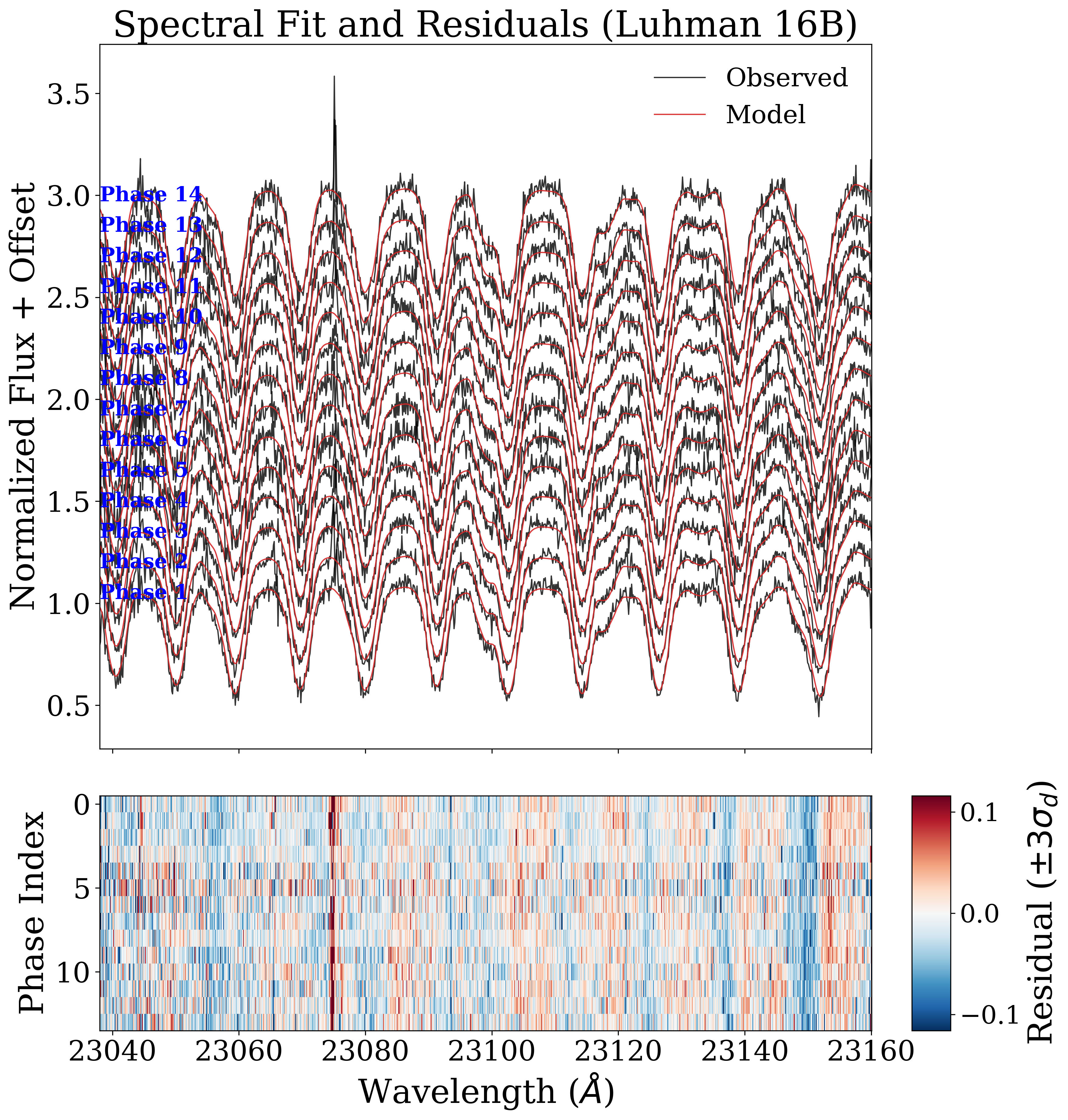}
  \caption{Spectral fit and residuals for Luhman 16B.
  \textit{Top panel:} Comparison between the observed spectral time series (black) and the model reconstruction based on the posterior median parameters (red). The spectra are vertically offset for clarity.
  \textit{Bottom panel:} Two-dimensional map of the residuals (observation minus model) normalized by the inferred noise amplitude $\bar{\sigma_d}$. The color scale covers the range of $\pm 3\bar{\sigma_d}$.
  The residuals show no phase-dependent drifting structures, although a prominent detector artifact is visible at $\lambda \approx 23075$ \AA}
  \label{fig:combined_fit}
\end{figure*}

To assess model-data consistency, we compute the spectral time series implied by the posterior solution.
Figure~\ref{fig:combined_fit} compares the observed spectra of Luhman 16B with the model predictions evaluated at the posterior median nonlinear parameters (Table~\ref{tab:luhman16b_params}) and the posterior mean surface map (Figure~\ref{fig:map_luh16B}). The model reproduces the time-dependent line profile distortions induced by rotation, capturing the evolution of the absorption features over a full rotation period.

The lower panel of Figure~\ref{fig:combined_fit} shows the residuals, $\boldsymbol{r}=\boldsymbol{d}-W(\bar{\boldsymbol{\theta}})\bar{\boldsymbol{\mu}}$. Their amplitude is broadly consistent with the inferred noise level ($\sigma_d \approx 0.039$), except for an outlier near $\lambda \approx 23075$ \AA, likely due to a telluric residual. The residual map shows no significant phase-dependent structure, indicating that the rotational variability is well captured by the inferred surface map. Some wavelength-dependent systematics remain across all phases, appearing as vertical stripes. Overall, the agreement supports the adequacy of the Bayesian Doppler imaging model for describing the spectral variability of Luhman 16B.

\section{Discussion}
\subsection{Surface Inhomogeneity and Comparison with Previous Work}
\begin{figure}[htbp]
  \centering
  \includegraphics[width=\linewidth]{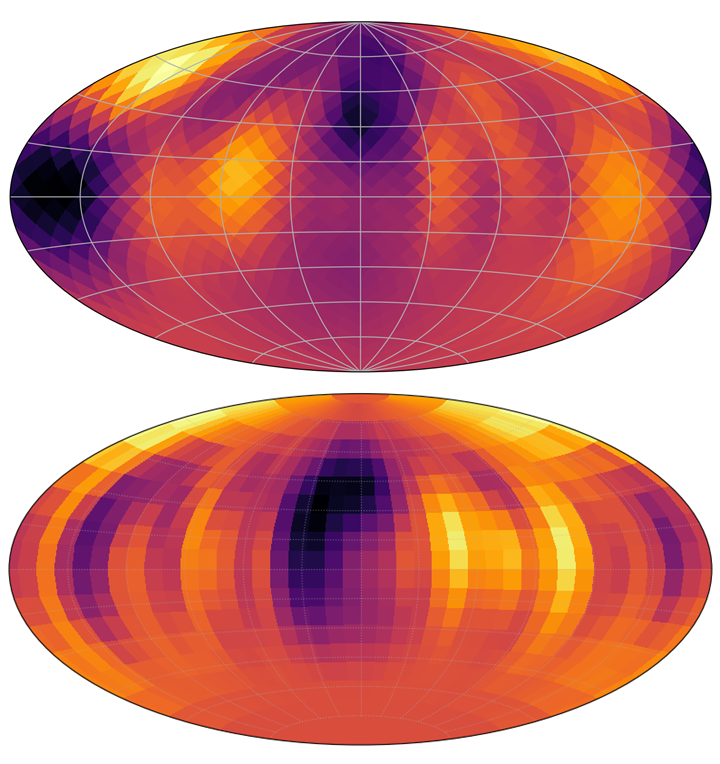}
  \caption{Comparison of surface maps. (Top) Our posterior mean surface map inferred from Luhman 16B data, shifted by $180^\circ$ in longitude relative to Fig.~\ref{fig:map_luh16B} and presented in the Aitoff projection. (Bottom) The surface map reconstructed by \citet{2014Natur.505..654C}.}
  \label{fig:map_comparison}
\end{figure}

Our Bayesian Doppler imaging analysis has revealed a significant surface inhomogeneity on Luhman 16B. 
The posterior mean map (Figure \ref{fig:map_comparison}, Top) exhibits a large,
mid-latitude dark region covering a wide range of longitudes,
together with several bright patches near high latitudes.

{
The large-scale dark region is qualitatively consistent with the dominant dark structure recovered by \citet{2014Natur.505..654C}  and with the overall morphology reported by \citet{2021arXiv211006271L}. This agreement supports the interpretation that the large dark feature is data-driven rather than solely an artifact of our particular inversion scheme. However, based on the synthetic tests, we caution that isolated bright patches, especially at high latitudes, may be affected by latitudinal degeneracy and by the GP prior, and should therefore be regarded as tentative.
Additionally, while our analysis and \citet{Chen_2024} targeted different epochs, the typical size of the surface inhomogeneities we recovered is comparable to the scale reported in their work, although a precise comparison is somewhat limited by the inherent latitudinal elongation of spots in Doppler imaging.
}

\begin{figure}[htbp]
  \centering
  \includegraphics[width=\linewidth]{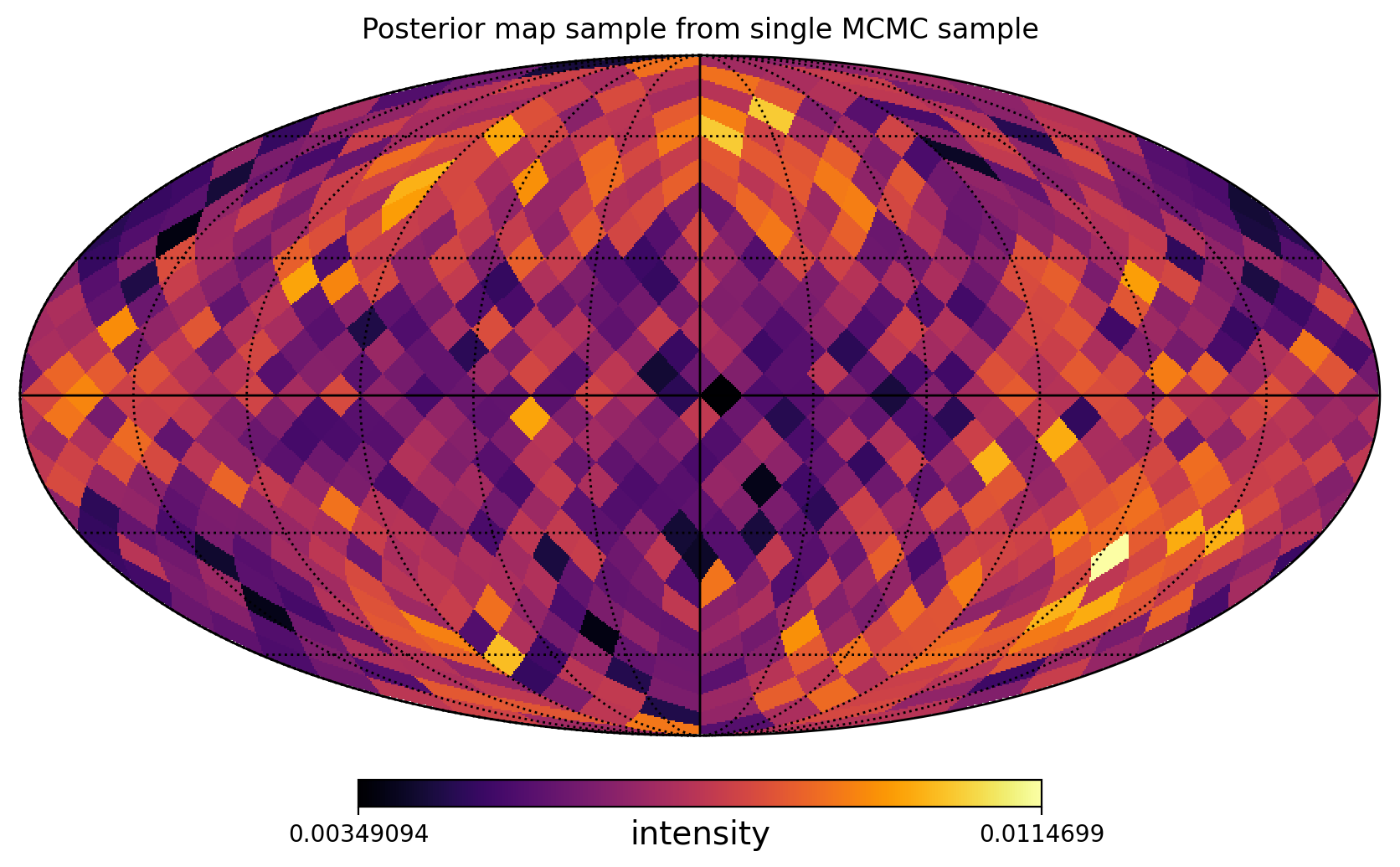}
  \caption{A representative sample of the surface intensity map drawn from the posterior distribution. While the dark region observed in the posterior mean map is also visible here, the map exhibits high-frequency granular noise. This pixel-scale fluctuation is partly attributable to the jitter term included in the covariance matrix for numerical stability.}
  \label{fig:individual_map_sample}
\end{figure}

A key advantage of the Bayesian approach is the ability to quantify uncertainties and assess the spatial coherence of the features. By inspecting individual samples (Figure~\ref{fig:individual_map_sample}) drawn from the posterior distribution of the surface map, we confirmed that the coherent dark region is present in individual realizations, rather than emerging solely as an artifact of averaging. Furthermore, the inferred correlation length, $\ell \approx 0.40$ rad (corresponding to $\approx 23^\circ$ on the sphere), provides a measure of the characteristic size of the surface features. This value is physically comparable to the angular scale of the recovered spots, indicating that the GP kernel successfully captures the typical size of the ``weather'' patterns on Luhman 16B without overfitting to pixel-scale noise. {These results reinforce the interpretation that the dominant dark 
inhomogeneity is associated with large-scale cloud structure affecting 
the flux variability, while the isolated bright patches remain tentative.}

\subsection{Geometric Properties and the Luhman 16 System}
The simultaneous inference of geometric parameters allows us to place constraints on the physical properties of Luhman 16B and the binary system configuration. We derived an equatorial rotation velocity of $v_{\mathrm{rot}} = 31.2_{-3.1}^{+5.3}~\mathrm{km\,s^{-1}}$ and an inclination of $i = {61.0}_{-12.3}^{+14.3}$ degrees. Crucially, our method {disentangles the usual} degeneracy between $v_{\mathrm{rot}}$ and $\sin i$ that typically limits standard line-broadening analysis. This disentanglement is achieved because the detailed, time-dependent distortions of the line profiles track surface features crossing specific latitudes, providing geometric constraints on the inclination independent of the total broadening width.

{In previous Doppler-imaging analyses of Luhman 16B (Table \ref{tab:comparison}), $v_{\mathrm{rot}}$ and $i$ were generally not inferred jointly within the surface-mapping likelihood. \citet{2014Natur.505..654C} measured $v_{\mathrm{rot}}\sin i = 26.1 \pm 0.2~\mathrm{km\,s^{-1}}$ and argued for a nearly equator-on viewing geometry, but noted that the inclination 
could not be directly constrained by the Doppler-imaging data. 
Subsequent reanalyses of the same CRIRES data by \citet{2021arXiv211006271L} 
and \citet{Plummer_2022} adopted a fiducial inclination of 
$i=70^\circ$ and used the Crossfield et al. value, either as a tight prior 
or as a fixed value.  \citet{Chen_2024} measured 
$v_{\mathrm{rot}}\sin i \simeq 27.2$--$27.5~\mathrm{km\,s^{-1}}$ from 
Gemini/IGRINS spectra and adopted $i=80^\circ$ for Luhman 16B Doppler maps. In contrast, our Bayesian framework treats $v_{\mathrm{rot}}$ and $i$ as nonlinear parameters in the Doppler-imaging 
likelihood and constrains them jointly from the spectral time series, 
without fixing either parameter to literature values. We infer 
$v_{\mathrm{rot}}=31.2^{+5.3}_{-3.1}~\mathrm{km\,s^{-1}}$ and 
$i=61.0^{+14.3}_{-12.3}$ deg, corresponding to 
$v_{\mathrm{rot}}\sin i \simeq 27.3~\mathrm{km\,s^{-1}}$, broadly 
consistent with previous projected rotational velocity estimates.}

\begin{table*}[htbp]
\centering
\caption{Comparison of projected rotational velocities and inclination assumptions/inferences for Doppler-imaging studies of Luhman 16B. 
Values listed for previous studies are not homogeneous posterior constraints: in most cases, $v_{\mathrm{rot}}\sin i$ was measured or adopted prior to map reconstruction, while $i$ was fixed or estimated using auxiliary information.}
\begin{tabular}{llll}
\hline\hline
Study & Data & $v_{\mathrm{rot}}\sin i$ treatment & Inclination treatment\\
\hline
\citet{2014Natur.505..654C} 
& VLT/CRIRES, K band 
& measured $26.1\pm0.2~\mathrm{km\,s^{-1}}$ 
& not directly inferred$^\dagger$ \\
%& maximum-entropy Doppler map \\

\citet{2021arXiv211006271L} 
& Crossfield et al. CRIRES data 
& tight prior $\mathcal{N}(26.1,0.2)~\mathrm{km\,s^{-1}}$ 
& fixed $i=70^\circ$ \\
%& spherical-harmonic \texttt{starry} MAP reconstruction \\

\citet{Plummer_2022} 
& Crossfield et al. CRIRES data 
& fixed $26.1~\mathrm{km\,s^{-1}}$ 
& fixed $i=70^\circ$ \\
%& analytical spot model; nested sampling for spot parameters \\ 

\citet{Chen_2024} 
& Gemini/IGRINS H+K bands 
& measured $27.2\pm4.1$/$27.5\pm4.3~\mathrm{km\,s^{-1}}$ 
& adopted $i=80^\circ$ \\
%& maximum-entropy H/K-band maps \\

This work 
& VLT/CRIRES, K-band chip 2 
& inferred $v_{\mathrm{rot}}\sin i \simeq 27.3~\mathrm{km\,s^{-1}}$ 
& inferred $i=61^{+14}_{-12}$ deg \\ 
%& $v_{\rm rot}$ and $i$ jointly inferred \\
\hline
\end{tabular}
\tablecomments{$\dagger$; near-equator-on / fiducial $i\simeq70^\circ$ in subsequent comparisons }
\label{tab:comparison}
\end{table*}

We also explicitly inferred the rotation period $P$ as a nuisance parameter to ensure the phase consistency of the surface reconstruction. Our analysis yielded a posterior estimate of $P = 4.828_{-0.018}^{+0.042}$\,hr. 
It should be noted that the joint estimation of $P$ is somewhat unstable and sensitive to the setup. In some cases, poorly defined initial conditions lead to anomalous results, including $v_{\mathrm{rot}}$ hitting the prior's upper limit instead of reaching a true global minimum.
Although the observation duration ($\sim 5$\,hr) covers only one full rotation and thus lacks the multi-rotation baseline typically required for precise period determination, our results show that $P$ is constrained within a narrow range around a value closer to the photometric period reported by \citet{2013A&A...555L...5G} ($4.87$\,hr) rather than the longer period derived from multi-epoch observations by \citet{2021ApJ...906...64A} ($5.28$\,hr). Given the possibility of differential rotation on brown dwarfs \citep[e.g.,][]{2021ApJ...906...64A}, our inferred period likely reflects the rotation rate of the latitudes where the dominant surface features (i.e., the large dark region and the spot in Figure \ref{fig:map_luh16B}) are located. Furthermore, we confirmed that fixing $P$ to literature values did not lead to significant changes in the estimation of other parameters or the surface map compared to the results obtained by treating $P$ as a free parameter (e.g., all key parameters remained consistent to two significant figures for $P=5.0\,\mathrm{hr}$).

\begin{figure}[htbp]
  \centering
  \includegraphics[width=\linewidth]{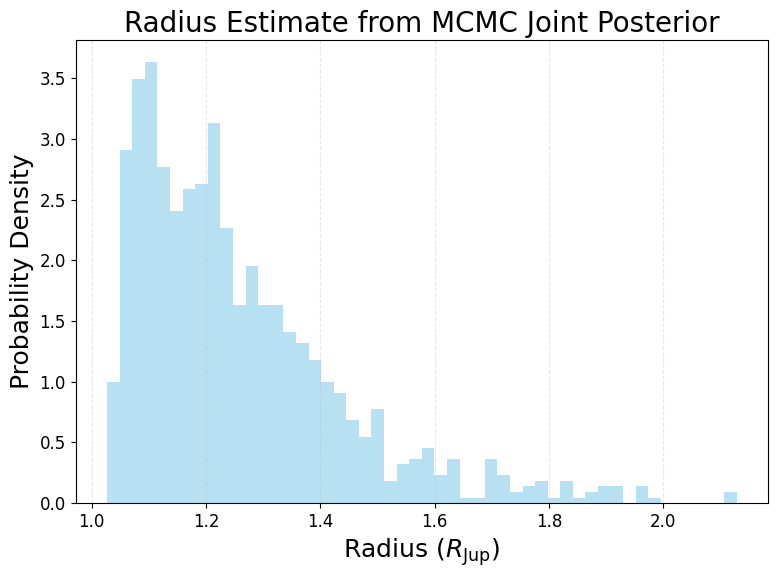}
  \caption{Posterior probability distribution of the radius of Luhman 16B. The radius is derived from the joint posterior samples of the equatorial rotation velocity $v_{\mathrm{rot}}$ and the rotation period $P$ using the relation $R = v_{\mathrm{rot}} P / (2\pi)$.}
  \label{fig_radius}
\end{figure}

Combining our inferred equatorial rotation velocity ($v_{\mathrm{rot}} \approx 31.2\,\mathrm{km\,s^{-1}}$) with the period ($P \approx 4.83$\,hr), we estimate the radius of the brown dwarf to be $R = v_{\mathrm{rot}} P / (2\pi) \approx 1.2\,R_{\mathrm{Jup}}$. Although this median value is slightly larger than the radius predicted by evolutionary models for an old brown dwarf ($\sim 0.9$--$1.0\,R_{\mathrm{Jup}}$; e.g., \citealt{2003A&A...402..701B}), the posterior distribution of the radius (Figure~\ref{fig_radius}) exhibits a tail extending down to $\approx 1.0\,R_{\mathrm{Jup}}$. Therefore, considering the uncertainties, our result remains broadly consistent with the theoretical predictions.

Regarding the limb darkening, our inferred quadratic coefficients ($q_1 = 0.81_{-0.22}^{+0.13}$, $q_2 = 0.59_{-0.13}^{+0.16}$) are broadly consistent with those derived from the atmospheric retrieval by \citet{2025arXiv251123018Y} ($q_1 = 0.80_{-0.09}^{+0.19}$, $q_2 = 0.34_{-0.12}^{+0.11}$). 

Comparing our results with the primary component, Luhman 16A, offers an interesting insight into the formation of the binary system. More broadly, Brown dwarf binaries, including recently resolved systems such as Gl 229Ba/Bb \citep{2024Natur.634.1070X, 2025ApJ...988...53K}, provide important constraints on substellar formation and angular momentum evolution. Luhman 16A is known to have a significantly lower projected rotational velocity of $v_{\mathrm{rot}} \sin i \approx 17.6\,\mathrm{km\,s^{-1}}$ \citep{2014Natur.505..654C}, compared to $\approx 26.1\,\mathrm{km\,s^{-1}}$ (literature value) or our derived $\approx 27.3\,\mathrm{km\,s^{-1}}$ for Luhman 16B. Assuming that the two components of the binary formed from the same cloud and share similar equatorial rotation velocities ($v_{\mathrm{rot}} \approx 31.2\,\mathrm{km\,s^{-1}}$), the observed difference in line broadening implies a difference in inclination. If Luhman 16A rotates at the same speed as B, its inclination would be constrained to $i_A \approx \arcsin(17.6/31.2) \approx 34^\circ$. Given our posterior estimate of $i_B \approx 61^\circ$, this would suggest a significant misalignment between the spin axes of the two brown dwarfs. Although precise measurements of $v_{\mathrm{rot}}$ for Luhman 16A are required to confirm this hypothesis, our method demonstrates the potential of Bayesian Doppler imaging to probe the spin-orbit architecture of substellar binaries.

\subsection{Methodological Limitations and Future Prospects}

While the overall spectral variability is well reproduced, the residual map (Figure \ref{fig:combined_fit}, bottom) reveals systematic vertical stripes that persist across all rotational phases. Close inspection of these residuals suggests that, in some spectral lines, the discrepancy exhibits an antisymmetric pattern around the line center (i.e., S-shaped residuals). This characteristic shape strongly implies a slight mismatch in the system radial velocity (RV). In this work, we fixed the intrinsic line profile $\boldsymbol{s}^\ast$ to the best-fit spectrum obtained from the atmospheric retrieval by \citet{2025arXiv251123018Y}, which already included a specific RV shift. However, since the optimal RV is generally covariant with other line-broadening parameters such as $v_{\mathrm{rot}}\sin i$ and limb darkening coefficients, the RV values derived in the retrieval are intrinsically coupled with the specific estimates of these broadening parameters obtained in that study. Therefore, utilizing these fixed RV values may result in a slight inconsistency when paired with the updated geometric parameters inferred in our fully Bayesian Doppler imaging. Even a minor velocity offset, when fixed in $\boldsymbol{s}^\ast$, would manifest as non-rotating, vertical stripes in the residual map.

Another technical consideration involves the numerical stability of the Gaussian Process. To perform the Cholesky decomposition of the covariance matrix $\Sigma_a + W^\mathsf{T} \Sigma_d^{-1} W$ (or its Woodbury equivalent) stably, a small ``jitter'' term must be added to the diagonal elements. We found that a jitter value of approximately $5\times10^{-7}$ was necessary to ensure convergence. While this stabilization is standard practice, there is a trade-off: an excessively large jitter ($\gtrapprox 10^{-6}$) can result in noisy posterior map samples (Figure~\ref{fig:individual_map_sample}), while too small a value leads to numerical instability. This implies a practical limit to the sensitivity of the method to extremely small-scale or low-contrast surface features.

To address the issue of the intrinsic profile mismatch, a promising future direction is the simultaneous inference of the surface map and the intrinsic spectrum. This can be achieved by including the atmospheric parameters governing $\boldsymbol{s}^\ast$ (e.g., coefficients of the T-P profile) directly into the set of nonlinear parameters $\boldsymbol{\theta}$ for joint sampling. Alternatively, as demonstrated in Appendix \ref{app:bilinearity}, the bilinearity of the forward model offers a pathway to infer the intrinsic spectrum iteratively or non-parametrically. Extending our Bayesian framework to incorporate these strategies would effectively mitigate the systematic residuals and provide a more self-consistent view of the brown dwarf's atmosphere.

\section{Summary and Conclusions}
\label{sec:conclusion}

We have presented a fully Bayesian framework for Doppler imaging that jointly infers surface brightness maps and geometric parameters from high-resolution spectral time series. Modeling the surface map as a Gaussian Process and analytically marginalizing the pixel intensities reduces the inference to the nonlinear parameters, which are sampled efficiently with Hamiltonian Monte Carlo. The method provides quantitative uncertainties for both surface structure and global geometry.

{Tests with synthetic data show that the framework recovers the 
longitudes of large-scale dark inhomogeneities and constrains 
\((v_{\mathrm{rot}}, i)\) under the adopted model assumptions, while 
also revealing limited sensitivity to latitude and isolated bright patches.}
Applied to VLT/CRIRES observations of Luhman 16B, we infer $i = 61.0_{-12.3}^{+14.3}$ degrees and $v_{\mathrm{rot}} = 31.2_{-3.1}^{+5.3}~\mathrm{km,s^{-1}}$, implying a radius consistent with evolutionary models of brown dwarfs. The reconstructed map reveals a large-scale dark region at mid-latitudes, in agreement with previous studies, now with spatially resolved uncertainty estimates.

By treating the surface map as a latent variable, the framework constrains both atmospheric structure and fundamental parameters within a unified probabilistic model. Extensions such as joint inference of intrinsic spectral profiles may further reduce systematic residuals and broaden the applicability of Doppler imaging to other substellar and exoplanetary atmospheres.

\section*{Acknowledgments}

We gratefully acknowledge Ian J. M. Crossfield for providing the reduced time-series spectra of Luhman 16B and the surface map data used in \cite{2014Natur.505..654C}.
We thank Shota Miyazaki, Yui Kasagi, Shotaro Tada, Ko Hosokawa, Yui Kawashima, and Benjamin Pope for the fruitful discussions. {We also thank an anonymous reviewer for the constructive comments.}
This study was supported by JSPS KAKENHI grant nos. 21H04998, 23H00133, 23H01224, 26H02072, 26K00752, and 26H02074 (H.K.).
We used ChatGPT (OpenAI) for language editing. No part of the scientific analysis was performed by ChatGPT. The authors are responsible for the content and interpretation.

\software{ ExoJAX \citep{Kawahara_2022, Kawahara_2025}, matplotlib, numpy, JAX \citep{jax2018github}, NumPyro \citep{phan2019composable}, Healpy \citep{2005ApJ...622..759G, Zonca2019}, healjax \citep{healjax}, corner \citep{Foreman-Mackey2016}}

\clearpage
\appendix
\section{Radial velocity of Surface Elements}
\label{app:rv_ele}
%%%
Each surface element is first expressed in Cartesian coordinates in the stellar frame as
\begin{align}
  \begin{pmatrix}
    x^\ast \\ y^\ast \\ z^\ast
  \end{pmatrix}
  =
\begin{pmatrix}
    \sin\theta^\ast_j \cos\phi^\ast_{jk} \\ \sin\theta^\ast_j \sin\phi^\ast_{jk} \\ \cos\theta^\ast_j
\end{pmatrix}
.
%  x^\ast &= \sin\theta^\ast_j \cos\phi^\ast_{jk}, \\
%  y^\ast &= \sin\theta^\ast_j \sin\phi^\ast_{jk}, \\
%  z^\ast &= \cos\theta^\ast_j.
\end{align}
The coordinates are then rotated about the $y^\ast$-axis by an angle $\alpha = \pi/2 - i$, where $i$ is the inclination of the rotation axis:
\begin{align}
  \begin{pmatrix}
    x \\ y \\ z
  \end{pmatrix}
  &= R_y(\alpha)
  \begin{pmatrix}
    x^\ast \\ y^\ast \\ z^\ast
  \end{pmatrix}
  =
  \begin{pmatrix}
    \cos\alpha & 0 & \sin\alpha \\
    0 & 1 & 0 \\
    -\sin\alpha & 0 & \cos\alpha
  \end{pmatrix}
  \begin{pmatrix}
    x^\ast \\ y^\ast \\ z^\ast
  \end{pmatrix}.
\end{align}
Finally, transforming back to spherical coordinates yields the apparent coordinates $(\theta_{jk}, \phi_{jk})$ in the observer’s frame: \begin{align}
  \theta_{jk} &= \arccos z, \\
  \phi_{jk} &= \arctan2(y, x).
\end{align}
In this coordinate system, $(\theta,\phi)=(\alpha,0)$ corresponds to the north pole, and $(\theta,\phi)=(\pi/2,0)$ points toward the observer (see Figure~\ref{fig1}).

Each surface element possesses a velocity due to the spin of the object.  Let $v_{\mathrm{rot}}$ denote the equatorial rotation velocity.  When the spin axis is perpendicular to the line of sight ($\alpha = 0$), the velocity vector in the co-rotating frame is
\begin{align}
  \boldsymbol{v^\ast}
  = v_{\mathrm{rot}}\sin\theta^\ast_j
    \begin{pmatrix}
      -\sin\phi^\ast_{jk} \\[3pt]
       \cos\phi^\ast_{jk} \\[3pt]
       0
    \end{pmatrix}.
\end{align}
If the axis is tilted, the vector is rotated about the $y^\ast$-axis by
$\alpha = \pi/2 - i$:
\begin{align}
  \boldsymbol{v}
  &= R_y(\alpha)\,\boldsymbol{v^\ast} \\
  &= v_{\mathrm{rot}}\sin\theta^\ast_j
    \begin{pmatrix}
      -\cos\alpha\sin\phi^\ast_{jk} \\[3pt]
       \cos\phi^\ast_{jk} \\[3pt]
       \sin\alpha\sin\phi^\ast_{jk}
    \end{pmatrix}.
\end{align}

The line-of-sight component of the velocity is obtained by taking the inner product with the unit vector pointing from the observer toward the object,
$\hat{\boldsymbol{n}}_{\mathrm{obs}} = (-1,0,0)$:
\begin{align}
  v_{\mathrm{los}}
  &= \hat{\boldsymbol{n}}_{\mathrm{obs}}\!\cdot\!\boldsymbol{v} \\
  &= v_{\mathrm{rot}}\cos\alpha\,
     \sin\theta^\ast_j\sin\phi^\ast_{jk}.
\end{align}
%%%

\section{Doppler-shift operation by the linear interpolation}
\label{app:interpolation}

\begin{figure}[htbp]
  \centering
  \includegraphics[width=0.5\linewidth]{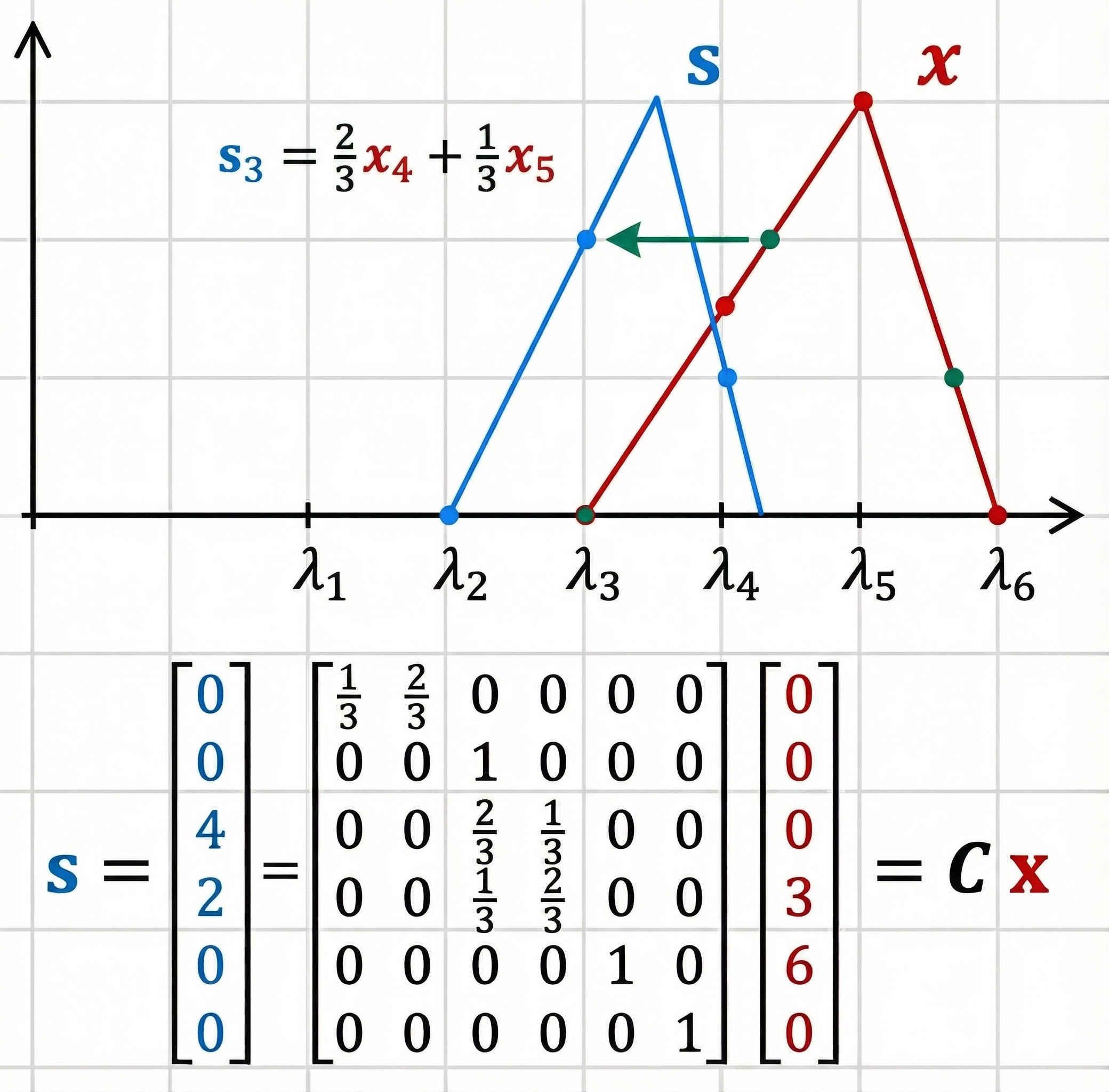}
  \caption{Example of linear interpolation for
  $N_{\mathrm{wav}}=6$ and $D=3/4$.}
  \label{fig2}
\end{figure}

{Figure~\ref{fig2} shows a schematic picture of the wavelength interpolation we apply.} This interpolation can be written using an $N_{\mathrm{wav}}\!\times\! N_{\mathrm{wav}}$ coefficient matrix $C^{(jk)}$ as
\begin{align}
  \boldsymbol{s}_{jk} = C^{(jk)}\,\boldsymbol{s}^\ast.
\end{align}
The $(l,l')$ element of $C^{(jk)}$, denoted $c^{(jk)}_{ll'}$, is constructed as follows. For each wavelength grid point $\lambda_l\ (l=1,2,\ldots,N_{\mathrm{wav}})$ with spacing $\Delta\lambda$, define
\begin{align}
  \ell^{(jk)}_{l}
  = \frac{\lambda_l/D_{jk}-\lambda_1}{\Delta\lambda} + 1.
\end{align}
The index is then clipped to the valid range
$1 \le \hat{\ell}^{(jk)}_{l} \le N_{\mathrm{wav}}$ by
\begin{align}
  \hat{\ell}^{(jk)}_{l}
  = \max\!\big(1,\min(\ell^{(jk)}_{l},N_{\mathrm{wav}})\big),
\end{align}
and the interpolation coefficient is given by
\begin{align}
  c^{(jk)}_{ll'}
  = \max\!\big(0,\,1 - |\hat{\ell}^{(jk)}_{l}-l'|\big).
\end{align}

\section{Explicit form of the conditional Gaussian distribution}
\label{ap:cond_gauss}
Starting from Bayes' theorem,
\begin{align}
  p(\boldsymbol{a}\mid\boldsymbol{d},\boldsymbol{\theta})
  \propto
  p(\boldsymbol{d}\mid\boldsymbol{a},\boldsymbol{\theta})\,
  p(\boldsymbol{a}\mid\boldsymbol{\theta}),
\end{align}
we write the log-posterior as
\begin{align}
  \log p(\boldsymbol{a}\mid\boldsymbol{d},\boldsymbol{\theta})
  &= -\frac12
     (\boldsymbol{d}-W\boldsymbol{a})^\mathsf{T}
     \Sigma_d^{-1}
     (\boldsymbol{d}-W\boldsymbol{a})
     \nonumber\\
  &\phantom{=}~
     -\frac12
     (\boldsymbol{a}-\boldsymbol{\mu}_a)^\mathsf{T
     }\Sigma_a^{-1}
     (\boldsymbol{a}-\boldsymbol{\mu}_a)
     + \mathrm{const}.
\end{align}
Expanding the quadratic forms and collecting the terms that depend on $\boldsymbol{a}$ gives
\begin{align}
  \log p(\boldsymbol{a}\mid\boldsymbol{d},\boldsymbol{\theta})
  &= -\tfrac12\,\boldsymbol{a}^\mathsf{T}
       \bigl(\Sigma_a^{-1}+W^\mathsf{T}\Sigma_d^{-1}W\bigr)
       \boldsymbol{a}
     \nonumber \\
  &\phantom{=}~
     + \boldsymbol{a}^\mathsf{T}
       \bigl(W^\mathsf{T}\Sigma_d^{-1}\boldsymbol{d}
             + \Sigma_a^{-1}\boldsymbol{\mu}_a\bigr)
     + \mathrm{const}.
\end{align}
\section{Derivation of the Woodbury form of the posterior}
\label{app:woodbury}
For completeness, we summarize the derivation of Eqs.~(\ref{eq:Woodburied-covariance})-(\ref{eq:Woodburied-mean}). Starting from the posterior covariance in Eq.~(\ref{eq:posterior-covariance}), we apply the Woodbury identity,
\begin{align}
    (A+UCV)^{-1}=A^{-1}-A^{-1}U(C^{-1}+VA^{-1}U)^{-1}VA^{-1},\label{eq:Woodbury}
\end{align}
using the substitutions
\begin{align}
    A=\Sigma_a^{-1},\qquad U=W^\mathsf{T},\qquad C=\Sigma_d^{-1},\qquad V=W.
\end{align}
This gives
\begin{align}
    \Sigma_{a\mid d}=\Sigma_a-\Sigma_aW^\mathsf{T}(\Sigma_d+W\Sigma_aW^\mathsf{T})^{-1}W\Sigma_a.
\end{align}
For the posterior mean, Eq.~(\ref{eq:posterior-mean}) can be rewritten as
\begin{align}
    \boldsymbol{\mu}_{a\mid d}&=\bigl[\Sigma_a-\Sigma_aW^\mathsf{T}(\Sigma_d+W\Sigma_aW^\mathsf{T})^{-1}W\Sigma_a\bigr](\Sigma_a^{-1}\boldsymbol{\mu}_a+W^\mathsf{T}\Sigma_d^{-1}\boldsymbol{d})\\
    &=\boldsymbol{\mu}_a-\Sigma_aW^\mathsf{T}(\Sigma_d+W\Sigma_aW^\mathsf{T})^{-1}W\boldsymbol{\mu}_a+\Sigma_aW^\mathsf{T}\bigl[\Sigma_d^{-1}-(\Sigma_d+W\Sigma_aW^\mathsf{T})^{-1}W\Sigma_aW^\mathsf{T}\Sigma_d^{-1}\bigr]\boldsymbol{d}.
\end{align}
The bracketed term can again be rewritten using the Woodbury identity in Eq~(\ref{eq:Woodbury}) with
\begin{align}
    A=\Sigma_d,\qquad U=\Sigma_d,\qquad C=\Sigma_d^{-1},\qquad V=W\Sigma_aW^\mathsf{T},
\end{align}
and we obtain
\begin{align}
    \boldsymbol{\mu}_{a\mid d}&=\boldsymbol{\mu}_a-\Sigma_aW^\mathsf{T}(\Sigma_d+W\Sigma_aW^\mathsf{T})^{-1}W\boldsymbol{\mu}_a+\Sigma_aW^\mathsf{T}(\Sigma_d+W\Sigma_aW^\mathsf{T})^{-1}\boldsymbol{d}\\
    &=\boldsymbol{\mu}_a+\Sigma_aW^\mathsf{T}(\Sigma_d+W\Sigma_aW^\mathsf{T})^{-1}(\boldsymbol{d}-W\boldsymbol{\mu}_a).
\end{align}

\section{Bilinearity of the forward model}
\label{app:bilinearity}

In Section~\ref{construction_W}, the forward model at rotational phase $\varphi_k$ was written as a linear function of the surface map $\boldsymbol{a}$, \begin{align}
  \boldsymbol{d}_k
  &= \sum_{j=1}^{N_{\mathrm{pix}}}
     \max(0,\mu_{jk})\,L(\mu_{jk})\,\boldsymbol{s}_{jk}\,a_j,
  \label{eq:app-dk-sum}
\end{align}
where $\boldsymbol{s}_{jk}$ is the Doppler-shifted spectrum from pixel $j$ at phase $k$. Here we show that the same model is in fact bilinear in the surface map $\boldsymbol{a}$ and the intrinsic line profile $\boldsymbol{s}^\ast$.

Recall that the Doppler-shifted spectrum is obtained through the interpolation matrix $C^{(jk)}$:
\begin{align}
  \boldsymbol{s}_{jk} = C^{(jk)}\,\boldsymbol{s}^\ast.
\end{align}
Substituting this relation into Eq.~(\ref{eq:app-dk-sum}) gives
\begin{align}
  \boldsymbol{d}_k
  &= \sum_{j=1}^{N_{\mathrm{pix}}}
     \max(0,\mu_{jk})\,L(\mu_{jk})\,
     C^{(jk)}\,\boldsymbol{s}^\ast\,a_j.
\end{align}

We now construct the block matrix $C^{(k)}\in\mathbb{R}^{N_{\mathrm{wav}} \times (N_{\mathrm{pix}}N_{\mathrm{wav}})}$ by horizontally concatenating the weighted interpolation matrices:
\begin{align}
  C^{(k)}
  = \bigl[
      \max(0,\mu_{1k})L(\mu_{1k})\, C^{(1k)}\;\;
      \cdots\;\;
      \max(0,\mu_{N_{\mathrm{pix}}k})L(\mu_{N_{\mathrm{pix}}k})\, C^{(N_{\mathrm{pix}}k)}
    \bigr].
\end{align}
Introduce the Kronecker product $\boldsymbol{a}\otimes\boldsymbol{s}^\ast\in \mathbb{R}^{N_{\mathrm{pix}}N_{\mathrm{wav}}}$, whose $j$-th block is $a_j\boldsymbol{s}^\ast$. Then
\begin{align}
  \boldsymbol{d}_k
  &= C^{(k)}\,(\boldsymbol{a}\otimes\boldsymbol{s}^\ast).
\end{align}

Stacking all phases and defining
\begin{align}
  C &=
  \begin{pmatrix}
    w_1 C^{(1)}\\
    w_2 C^{(2)}\\
    \vdots\\
    w_{N_{\mathrm{phase}}} C^{(N_{\mathrm{phase}})}
  \end{pmatrix},
\end{align}
the full data vector satisfies
\begin{align}
  \boldsymbol{d}
  &= C\,(\boldsymbol{a}\otimes\boldsymbol{s}^\ast).
  \label{eq:app-bilinear-main}
\end{align}
Thus the forward model is bilinear in $\boldsymbol{a}$ and $\boldsymbol{s}^\ast$.

Using the properties of the Kronecker product, we obtain two equivalent linear representations:
\begin{align}
  \boldsymbol{d}
  &= C\,(I_{N_{\mathrm{pix}}}\otimes\boldsymbol{s}^\ast)\,\boldsymbol{a},
  \label{senkei_a}\\
  &= C\,(\boldsymbol{a}\otimes I_{N_{\mathrm{wav}}})\,\boldsymbol{s}^\ast.
  \label{senkei_x}
\end{align}

Equation~(\ref{senkei_a}) expresses the model as linear in the surface map $\boldsymbol{a}$ for fixed $\boldsymbol{s}^\ast$. Importantly,
\begin{align}  
  W = C\,(I_{N_{\mathrm{pix}}}\otimes\boldsymbol{s}^\ast)
\end{align}
is exactly the design matrix used in the main text. Thus the familiar linear model $\boldsymbol{d}=W\boldsymbol{a}$ is recovered as a special case of the bilinear formulation.

The representation in Eq.~(\ref{senkei_x}) shows that, for fixed $\boldsymbol{a}$, the model is linear also in $\boldsymbol{s}^\ast$. In principle, this bilinear structure allows one to alternate between updating the surface map and updating the intrinsic spectrum, e.g.,
\begin{align}  
  \boldsymbol{a}^{(t+1)} \leftarrow 
   \arg\max_{\boldsymbol{a}} p(\boldsymbol{a}\mid \boldsymbol{d}, \boldsymbol{s}^{*(t)}), \qquad
  \boldsymbol{s}^{*(t+1)} \leftarrow
   \arg\max_{\boldsymbol{s}^\ast} p(\boldsymbol{s}^\ast \mid \boldsymbol{d}, \boldsymbol{a}^{(t)}),
\end{align}
or analogously in a Bayesian formulation.

We emphasize, however, that while such alternating updates are theoretically well-defined, their practical viability depends on the computational cost of constructing and applying the large block matrix $C$, as both $a$ and $s^\ast$ enter the forward operator.
For problems where $N_{\mathrm{pix}}$ or $N_{\mathrm{wav}}$ is large, these costs may become substantial. Nevertheless, the bilinear formulation clarifies that joint inference of the surface map and the intrinsic spectrum is conceptually possible and provides a foundation for future methodological extensions.

\vspace{\baselineskip}

\bibliography{sample63}{}
\bibliographystyle{aasjournal}
\end{document}